\documentclass[twocolumn,showpacs,preprintnumbers,amsmath,amssymb,superscriptaddress]{emulateapj}

\usepackage{graphicx}% Include figure files≈
\usepackage{dcolumn}% Align table columns on decimal point
\usepackage{bm}% bold math
\usepackage{multirow}
\usepackage{rotating}

\usepackage{aas_macros}

\newcommand\beq{\begin{equation}}
\newcommand\eeq{\end{equation}}
\newcommand\beqar{\begin{eqnarray}}
\newcommand\eeqar{\end{eqnarray}}

\usepackage{color}
\newcommand{\updated}[1]{#1}

\begin{document}

%\preprint{APS/123-QED}

\title{Grain Physics and IR Dust Emission in AGN Environments}% 

\author{Brandon S. Hensley}
\email{bhensley@astro.princeton.edu}
\affiliation{Department of Astrophysical Sciences,  Princeton
  University, Princeton, NJ 08544, USA}

\author{Jeremiah P. Ostriker}
\affiliation{Department of Astrophysical Sciences,  Princeton
  University, Princeton, NJ 08544, USA}
\affiliation{Department of Astronomy, Columbia University, New York, NY 10027, USA}

\author{Luca Ciotti}
\affiliation{Department of Physics and Astronomy, University of Bologna, via
  Ranzani 1, I-40127, Bologna, Italy}

\date{\today}% It is always \today, today,

\begin{abstract}
We study the effects of a \updated{detailed}
dust treatment on the properties and evolution of early-type galaxies
containing central black holes, as determined by AGN feedback. We find
that during cooling flow episodes, radiation pressure on the dust in and interior to infalling shells of cold gas can
greatly impact the amount of gas able to be accreted and therefore the
frequency of AGN bursts. However, the overall hydrodynamic evolution
of all models, including mass budget, is relatively robust to the
assumptions on dust. We find that IR
re-emission from hot dust can dominate the bolometric luminosity of the
galaxy during the early stages of an AGN burst, reaching values in
excess of $10^{46}$ erg/s. The AGN-emitted UV is largely absorbed, but
the optical depth in the IR does not exceed unity, so the
radiation momentum input never exceeds $L_{\rm BH}/c$. We constrain the viability of our models by comparing
the AGN duty cycle, \updated{broadband luminosities},
dust mass, black hole mass, and other model
predictions to current observations. These constraints force us to
models wherein the \updated{dust to metals ratios are $\simeq 1\%$ of
  the Galactic value}, and only
models with a dynamic dust to gas ratio are able to produce both quiescent
galaxies consistent with observations and high obscured fractions
during AGN ``on'' phases. During AGN outbursts, we predict that a
large fraction of the FIR luminosity can be attributed to warm
dust emission ($\simeq100$ K) from dense
dusty gas within $\leq 1$ kpc reradiating the AGN UV emission.
\end{abstract}


\maketitle

\section{Introduction}

It is well-established that the brightest active galactic nuclei (AGNs) and the most massive black
holes reside in giant elliptical galaxies \updated{\citep{Dunlop_2004}}. It is also accepted that the environment provided by
the galaxy, particularly the abundance of hot metal-rich gas, aging stars, and
heavily depleted dust, profoundly affects the supermassive black hole
(hereafter SMBH) evolution, and in turn the host system is modified by
radiative and mechanical feedback
\citep[][and references therein]{Begelman+Blandford+Rees_1984, Norman+Scoville_1988,
  Ciotti+Ostriker_1997,
  Ciotti+Ostriker_2012, Ostriker+Choi+Ciotti+etal_2010,
  Pellegrini+Ciotti+Ostriker_2012, Vogelsberger+etal_2013}. Indeed, the interplay between the
evolution of the black hole and the evolution of its host galaxy has been
frequently invoked in order to explain
the observed correlation between the properties of the black hole and
the galaxy \citep[e.g.][]{Springel+DiMatteo+Hernquist_2005, 
Ostriker+Ciotti_2005, Cattaneo+Faber+Binney+etal_2009,
Debuhr+Quataert+Ma_2011}. Putting the questions we seek to answer in
concrete terms: if the black hole in a galaxy
such as M87 were
to erupt as a quasar, what would we see? What mechanisms would govern
the interaction between the accreting gas of the galaxy and the erupting
black hole? What sources will contribute to the observed radiation?

A key ingredient in answering each of the above questions is the
presence of dust in the galaxy. The potential importance of dust can
be readily seen for several reasons. First, quasars emit at or near their
Eddington limit, which classically is set by the Thomson scattering opacity
of $0.4\ {\rm cm}^2/{\rm g}$. By comparison, the opacity of dust grains to
UV photons exceeds 1000 times this value, so the radiative forces on
dust can easily overwhelm other dynamical drivers during outbursts. Second, it is widely
accepted that a significant fraction of quasars are obscured. However,
the giant elliptical galaxies which host AGN typically have very small
optical depths, implying little dust. Therefore, not only does the
dust govern the observed radiation from the galaxy, but the dust
abundance itself may be a dynamic quantity that evolves in parallel
with quasar ``on'' and ``off'' phases. In this work, we seek to understand the
power of dust to influence the nature, evolution, and observational
signatures of giant elliptical galaxies by introducing the processes
that create and destroy dust within galaxies into our simulations.

One of the primary links between an AGN and its host galaxy is the
immense radiative output of the AGN. The radiative feedback mechanisms between the
central black hole and the accreting gas greatly influence the gas
dynamics and thus have been the focus of much study
\citep[][hereafter CO07]{Binney+Tabor_1995, Ciotti+Ostriker_1997, Thompson+Quataert+Murray_2005, Ciotti+Ostriker_2007}. Because dust grains efficiently absorb UV radiation and re-radiate in the
IR, the presence of grains can dramatically affect the gas
dynamics \citep[][and references therein]{Siebenmorgen+Heymann_2011}. In particular, infalling
shells of cold gas (a common feature at the onset of ISM cooling
episodes and detected in giant ellipticals \citep{Werner+etal_2013}) can be supported by radiation pressure, slowing
accretion, and altering the subsequent evolution. In turn, this
evolution is related to the so-called ``positive feedback'' mode in
which feedback from the AGN enhances star formation
\citep[CO07,][and references
therein]{Nayakshin+Zubovas_2012}. Observations have hinted at the presence of
radiation-supported cold gas around AGN \citep{Vasudevan+etal_2013}
though in the majority ($\simeq95$\%) of cases the gas is below its effective
Eddington limit \citep{Raimundo+etal_2010}. Nevertheless, the latter authors
note the inevitability of radiation-supported gas and its possible
role in AGN feedback. Radiation pressure on
dust grains may be instrumental in driving galactic winds
\citep{Coker+Thompson+Martini_2013}, though there is disagreement as
to whether this could be a substantial effect
\citep{Socrates+Sironi_2013}.

In a recent series of papers \citep{Ciotti+Ostriker+Proga_2009,
  Shin+Ostriker+Ciotti_2010, Ciotti+Ostriker+Proga_2010}, 1D
hydrodynamical simulations were used to study the mechanisms of AGN feedback.
The dust physics was implemented in a very simple,
phenomenological way. As sputtering effectively destroys dust in regions
of hot gas \citep{Draine+Salpeter_1979}, these previous papers approximated the sputtering
by reducing the dust to gas ratio in hot gas by roughly two orders of magnitude
relative to Galactic values in accord with observations of dust in elliptical galaxies. In
practice, a factor inversely proportional to the gas temperature
multiplies the fiducial absorption coefficient in the UV, optical, and IR.

In this paper, we improve significantly this aspect of the input physics. We seek to clarify the role of dust grains in AGN
feedback by developing and numerically implementing the basic
equations for dust production and destruction. \updated{While a
  similar but simpler version of the dust physics described here was implemented
  in the 2D simulations of \citet{Novak+Ostriker+Ciotti_2012}, we
  present a more generalized treatment and focus principally on the
  effects of different models of dust abundance on the hydrodynamical evolution of
  the simulation, both in its galaxy-scale properties and accretion
  physics. Additionally, we discuss the ability or inability of models
  to reproduce the observational properties of giant elliptical
  galaxies in the infrared during both quiescent and accretion phases.}

For the present hydrodynamical simulations we use an updated version of the 1D code
used in our previous studies, with the relevant modifications detailed
in Section~\ref{sec:simulations}. In particular, a major upgrade has
been made in the solution of the radiative transfer equation. We note
that in a recent related paper \citep{Novak+Ostriker+Ciotti_2012}
the new ``two-streams'' approach is also tested in the context of 2D
hydrodynamical simulations. 

We restrict our focus to the secular evolution of the
galaxy. While processes such as major and minor mergers or inflows of
cold gas from the IGM certainly occur and can account for a variety of
observations, we are most interested in which aspects of the evolution
of these galaxies can be accounted for by purely secular processes, i.e.
driven by well-understood stellar evolutionary processes. We
find that the AGN duty cycle, its time dependence, the black hole mass, and the IR luminosity are all naturally
explained with our purely secular model.

The paper is organized as follows: in Section~\ref{sec:physics} we describe
the dust grain physics employed in our simulations as well as a
discussion of two-stream radiative transfer. In
Section~\ref{sec:simulations} we describe the suite of simulations
conducted under different models for the grain abundance. In
Section~\ref{sec:survey}, we summarize the results of the simulations and
compare against observations to select the most viable models. We also
discuss the effects of dust
on the properties of AGN outbursts. We then
discuss the most viable models in the context of observations in
Section~\ref{sec:observations}, focusing on both
observational signatures predicted by our model as well as how current
observations constrain the survival, growth, and radiative properties
of dust in elliptical galaxies with AGN. We summarize the implications of these
comparisons in Section~\ref{sec:discussion}.

\section{Grain Physics}
\label{sec:physics}

In this Section we summarize the dust physics implemented in the new
version of the code and used for the hydrodynamic simulations.

To aid comparisons between models, we define the factor $D$ to
describe how much the dust is
depleted relative to what is observed in the Galaxy, i.e.

\begin{equation}
\label{eq:Ddef}
\ \mathrm{D} \equiv \frac{Z_{\rm MW}}{Z} \left(\frac{\rho_d}{\rho}\right) 
\left(\frac{\rho_d}{\rho}\right)_{\rm MW}^{-1}
~~~,
\end{equation}
where $\rho$ and $\rho_d$ are the gas and dust mass densities,
respectively, and $Z$ and $Z_{\rm MW}$ are the metallicities of the simulated galaxy
and the Milky Way, respectively. This is motivated by the fact that the
dust to gas ratio of a galaxy should scale linearly with the
metallicity. A depletion factor of 1 therefore indicates that the fraction of metals
incorporated in dust grains is the same as in the Milky Way. We adopt $Z_{\rm MW} = Z_\odot$, a Galactic dust to gas ratio of 0.01,
and $Z/Z_{\rm MW} = 4/3$ \updated{for all times in the simulations}.

\subsection{Grain Opacity}

Following CO07, we adopt the dust opacity values

\begin{equation}
\label{eq:opacity}
\ \kappa_\mathrm{Op} = 300 \updated{\frac{{\rm cm}^2}{\rm g}} \left(\frac{Z}{Z_{\rm MW}}\right) \times \mathrm{D}, \hspace{0.2cm} \kappa_\mathrm{UV} = 4
\kappa_\mathrm{Op}, \hspace{0.2cm} \kappa_\mathrm{IR} =
\frac{\kappa_\mathrm{Op}}{150} ,
\end{equation}
i.e. dependent on the dust to gas ratio in the galaxy relative to the
Milky Way. \updated{The bands correspond the bands used by
  \citet{Sazonov+Ostriker+Sunyaev_2004} to compute the broadband AGN
  output and are listed in Table~\ref{table:bands}.} Note that in
CO07,

\begin{equation}
\label{eq:co_dfactor}
\ \mathrm{D} = \frac{1}{1+T_4} ,
\end{equation}
where $T_4$ is the gas temperature in units of $10^4$ K and the
metallicity taken to be $Z_\odot$. For coronal
X-ray emitting gas in a typical elliptical galaxy, this corresponds to
a depletion factor of $\simeq 10^{-2}$. In the following, we present a
method for computing D in a much more physically consistent way than
Equation~\ref{eq:co_dfactor}.

\begin{table}
\label{table:bands}
\caption{Bandpasses}
\begin{center}
    \begin{tabular}{|l  l c|}
      \hline
    Band & Energy & AGN Output Fraction\\
    \hline 
X-ray & $E >$ 2 keV & 0.1 \\
UV & 13 $< E <$ 2 keV & 0.35 \\
Op & 1 $< E <$ 13 eV & 0.25 \\
IR & $E <$ 1 eV & 0.3 \\\hline
    \end{tabular}
    \\
    \end{center}
    The assumed energy limits for the bandpasses used
    in this work.
\end{table}

\subsection{Dust Abundance}
\label{subsec:continuity}

The primary source\updated{s} of dust in \updated{giant elliptical} 
galaxies \updated{include} the massive winds ejected by dying AGB stars during
thermal pulses (for example as planetary nebulae) and by supernova
explosions\updated{, as well as grain growth in the metal-rich
  ISM}. The ultimate fate of these winds is not fully understood,
but it is commonly accepted that some form of mixing and
thermalization with the preexisting ISM takes place on short
timescales \citep[e.g.][]{Bregman+Parriott_2009}. Therefore, it is natural to expect that the dust is
transported through the galaxy
by the large scale gas flows of early type galaxies \citep{Kim+Pellegrini_2012}, where it is sputtered by the hot ISM and, in cold gas, allowed to grow via collisions with metal
atoms. Moreover, when dust-bearing gas forms new stars, the dust is removed
from the ISM and sequestered into the stars. 

In the previous simulations discussed in the Introduction, these effects were addressed
qualitatively by assuming that the cold gas was grain-rich and the hot
gas was grain-poor. In practice, this was implemented by approximating
the dust to gas ratio at each radius by
Equation~(\ref{eq:co_dfactor}). However, this prescription
changes the dust abundance instantaneously with temperature
and does not allow for the transport of grains from one radius to
another.

\subsubsection{One Component Model}
\label{sec:one_component}

The new treatment addresses both issues by implementing a grain
continuity equation

\begin{equation}
\label{eq:continuity}
\ \frac{\partial \rho_{d}}{\partial t} + \nabla \cdot \left(\rho_d
  {\bf v}\right) = S_+ - S_- ,
\end{equation}
where $\rho_d$ is the mass density of dust grains. Note that {\bf v} is the \emph{gas}
velocity given by the hydrodynamic code since we assume that the grains are coupled to the motion
of the gas. \updated{Although drift relative to the gas will occur due
  to gravitational forces and anisotropic radiation fields,
  such drift will be mitigated by Coulomb drag and drag from the local magnetic field since the grains
  will be charged. The net drifts will in general be small relative to
  the Eulerian velocities.} Finally $S_+$ and $S_-$ are the dust source and sink terms,
respectively. As discussed below, each of the two functions $S_+$ and
$S_-$ is in turn given by the sum of two terms. We term this the ``One
Component'' model as all dust in this model is assumed to be mixed
with the ISM. We describe a ``Two Component'' model in
Section \ref{sec:two_component} that considers mixed and unmixed dust separately.

In the Milky Way, the dust to gas ratio in outflows of
oxygen-rich AGB stars has
been observed to be $\simeq 0.0063$ \citep{Knapp_1985, Kemper+Stark+Justtanont+etal_2003}. Since an early type galaxy can be
metal-enriched relative to the Milky Way, we adopt a fiducial dust to gas mass
ratio of 0.01 in these outflows, and the source term is simply

\begin{equation}
\ S_{+, inj} = 0.0133 \dot{\rho}_{*} ,
\end{equation}
where $ \dot{\rho}_{*}$ is the gas released by stars as described in
CO07 Equation~13, and the numerical factor
is dimensionless.

In the particularly hot
ISM of elliptical galaxies, grains are rapidly sputtered. To compute the dust
destruction rate due to sputtering, we use the relation

\begin{equation}
\ \dot{a} = -\frac{10^{-6} n_{\rm H}}{1 +
  T_6^{-3}}\ \mu\mathrm{m}\
\mathrm{yr}^{-1}
\end{equation}
where $a$ is the grain radius, $T_6$ is the gas temperature in
units of $10^6$ K, and $n_{\rm H}$ is the proton number density
in units of $\mathrm{cm}^{-3}$. This expression is a good approximation for graphite and silicate grains in gas
with $10^5 < T < 10^9$ K \citep{Draine_2011}; note that $\dot{a}$ is
independent of $a$. Empirically, the size distribution of dust grains
above $\simeq50\AA$ can be approximated by the standard Mathis-Rumpl-Nordsieck (MRN) distribution
\citep{Mathis+Rumpl+Nordsieck_1977}
despite destruction and creation processes. Therefore, we assume that
the MRN distribution is valid at all times, i.e.

\begin{equation}
\label{eq:mrn_dist}
\ \frac{\mathrm{d}n_d}{\mathrm{d}a} = \frac{H a_{\rm max}^{2.5}}{a^{3.5}}
\end{equation}
is the number density of grains with size
between $a$ and $a+$d$a$, where $H$ is a
normalization constant, in principle dependent on time and position in
the galaxy. In the following, we
assume $a_{\rm min} = 0.005 \mu$m and $a_{\rm max} = 0.3 \mu$m. The total
density in grains at a given radius and time is then

\begin{equation}
\label{eq:rho_gd}
\rho_{d} \simeq \frac{8 \pi}{3} H \rho_{\mathrm{grain}}
a_{\mathrm{max}}^3
~~~,
\end{equation}
where $\rho_{\mathrm{grain}}$ is the internal density of a dust grain
and we neglected the factor of $1 - \sqrt{a_{\mathrm{min}}/a_{\mathrm{max}}}$. We take
$\rho_{\mathrm{grain}} = 3.5$ g cm$^{-3}$ which is a standard value for silicate
grains. The total destruction rate is obtained by computing the mass destruction rate of grains
of radius $a$ and then integrating over the distribution. It follows that

\begin{equation}
\label{eq:rhodot_gd}
\ \dot{\rho}_{d,gd} = 8 \pi H \rho_{\rm grain}
\left(\frac{a_{\rm max}}{a_{\rm min}}\right)^{0.5}a_{\rm max}^2\dot{a} ,
\end{equation}
and from Eqs.~(\ref{eq:rho_gd})-(\ref{eq:rhodot_gd}), we can define
a grain destruction frequency

\begin{equation}
\label{eq:nugd}
\ \nu_{gd} \equiv \frac{|\dot{\rho}_{d,gd}|}{\rho_{d}} = 3
\left(\frac{a_{\mathrm{max}}}{a_{\rm min}}\right)^{0.5}
\frac{|\dot{a}|}{a_{\rm max}} .
\end{equation}
Therefore, the sputtering term in Equation~\ref{eq:continuity} is

\begin{equation}
\ S_{-,gd} = \nu_{gd} \rho_d .
\end{equation}

Note that in principle, in AGN environments, where high energy photons can ionize grains, the sputtering time can be altered by the effects
of grain charging. However,
\citet{Weingartner+Draine+Barr_2006} find that the effect at $r = 100\
$pc for an ${\rm L}_{\rm BH} = 10^{46}$ erg/s quasar is
negligible above $\simeq 10^6$ K, even for large ionization
parameters. Thus, this effect may be safely neglected.

In cold gas, metal atoms are able to collide with dust
grains and stick. If these metal atoms have a probability $f$ of
sticking and are moving with average speed $v_Z$, then the
source term due to these collisions is $f \rho_Z v_Z 4 \pi a^2 n_d$,
where $\rho_Z = f_Z
\rho$ is the mass density of the metals and $f_Z$ is the mass fraction
of gas available for making grains. If we assume that the Milky Way
has included all such materials in grains already, then $f_Z$ for the
Milky Way would just be its observed dust to gas ratio of 0.01. Since
the simulated galaxy has a metallicity $4/3$ greater than the Milky Way,
we take $f_Z$ to be $0.0133$. However, grain growth cannot continue after the metals in the gas have
been used up, so we replace $f_Z$ with $f_Z -
  \rho_{d}/\rho$ to disallow growth beyond the
available metal atoms. Integrating this over the grain size distribution, we obtain

\begin{equation}
\ \dot{\rho}_{d,gg} = f \rho \frac{c_s}{\sqrt{\mu_Z}} 8 \pi \left(f_Z -
  \frac{\rho_{d}}{\rho} \right) H a_{\rm max}^2
\left(\frac{a_{max}}{a_{\rm min}}\right)^{0.5} ,
\end{equation}
where we have made the approximation $v_Z = c_s/\sqrt{\mu_Z}$, with $c_s$
being the sound speed in the gas and $\mu_{\updated{Z}}$ the mean atomic mass of
the metal atoms. Thus the grain growth frequency is

\begin{equation}
\label{eq:nugg}
\ \nu_{gg} \equiv \frac{\dot{\rho}_{d,gg}}{\rho_d} = \frac{3 f \rho
  c_s}{\rho_{\rm grain} a_{\rm max}\sqrt{\mu_Z}} \left(f_Z -
  \frac{\rho_d}{\rho} \right)
\left(\frac{a_{\rm max}}{a_{\rm min}}\right)^{0.5} .
\end{equation}
\updated{Since grains are made primarily from carbon,
  oxygen, magnesium, silicon, and iron, we adopt $\mu_Z = 16$
  corresponding to oxygen and in agreement with the mean atomic mass
  in current grain models \citep[][Table 23.1]{Draine_2011}. Following
  \citet{Clayton_1976}, who modeled grain growth in
  the $10^4$ K gas of an expanding nova shell, we take $f =
  0.2$, but note that there is
  considerable uncertainty in the surface chemistry of grains,
  particularly in the high temperature environment of an elliptical
  galaxy}. 

\begin{figure}[htp]
 \centering
  		\scalebox{0.40}{\includegraphics{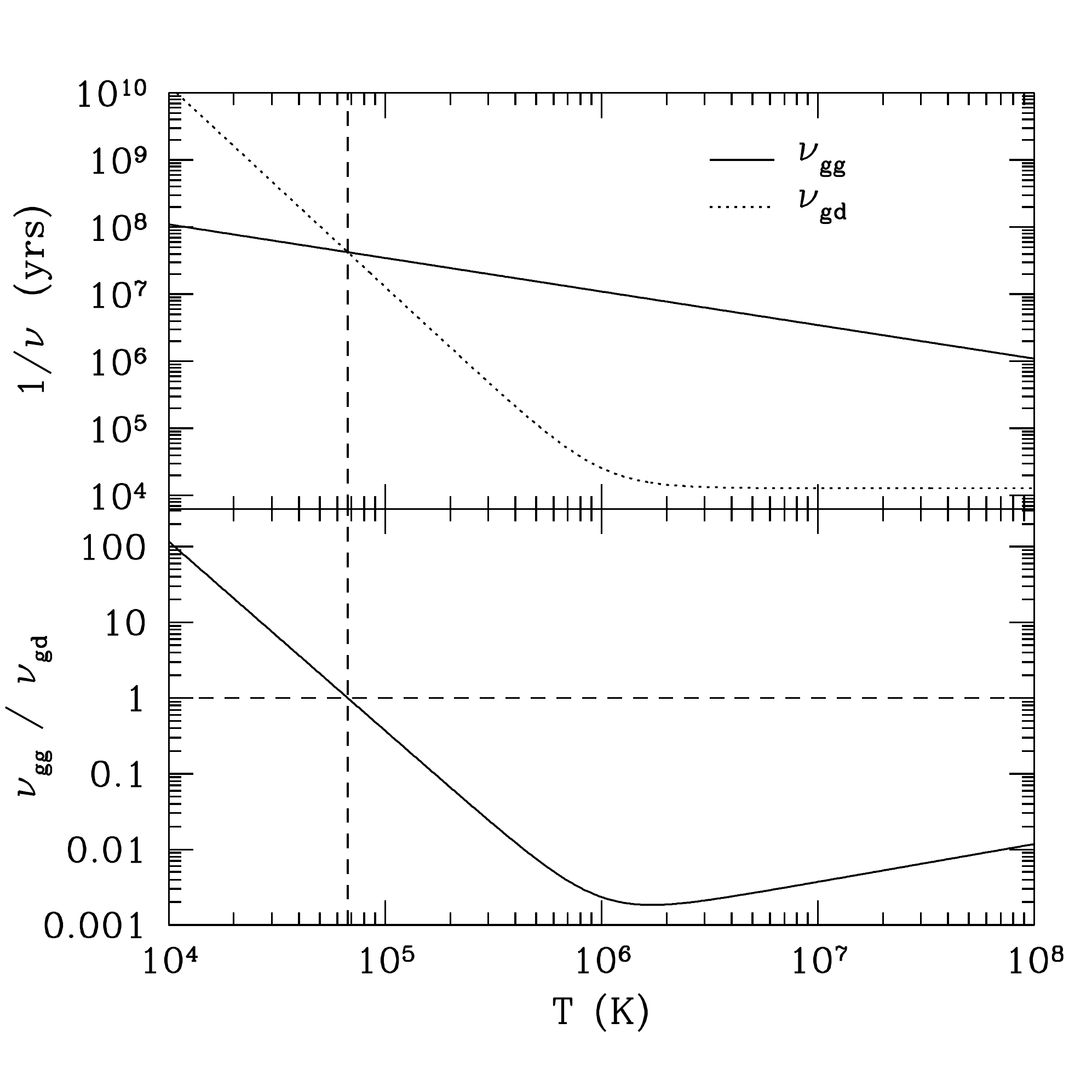}}
\caption{\emph{Top}: Grain growth (Equation~\ref{eq:nugg}) and
  destruction (Equation~\ref{eq:nugd}) times for a representative value $n_{\rm H} = 0.1$
  cm$^{-3}$ and $f_Z >>
\rho_d/\rho$. Net grain
  growth will only occur in gas colder than $\simeq 7 \times 10^4$
  K. Note that for these temperatures, grain growth time can be as
  short as $10^4$ years for a density of $10^3$ cm$^{-3}$ (see
Figure~\ref{fig:ce_rprof}) due to the linear dependence of the growth
frequency on density. \emph{Bottom}: The ratio of the growth to destruction
  frequencies (Equation~\ref{eq:nu_ratio}). } \label{fig:nu_ratio} 
\end{figure}

The ratio of grain growth by collisions to sputtering is a temperature
dependent function given by
\begin{equation}
\label{eq:nu_ratio}
\ \frac{\nu_{gg}}{\nu_{gd}} = 1.18\times10^{-6} \left(1 +
  T_6^{-3}\right) \sqrt{T}
\end{equation}
for our assumed galaxy parameters and approximating $f_Z >>
\rho_d/\rho$. Figure~\ref{fig:nu_ratio} plots the growth and
destruction times as a function of temperature for gas with $n_{\rm H} = 0.1$
  cm$^{-3}$. Grain growth is
negligible in all but the coldest gas.

%%%%%%%%%%%%%%%%%%%%%%%%%%%%%%%%%%%%%%%%%%%%%%%%%%%%%%%%%%%%%%%%%%%%%%%%%%%
% Figure-- Growth to Destruction Ratio
%%%%%%%%%%%%%%%%%%%%%%%%%%%%%%%%%%%%%%%%%%%%%%%%%%%%%%%%%%%%%%%%%%%%%%%%%%%

Finally, grains will be removed from the ISM when the gas containing
those grains form stars. The dust destruction due to
star formation is

\begin{equation}
\ S_{-,{\rm SF}} = \frac{\rho_d}{\rho_g} \dot{\rho}_*^+
~~~,
\end{equation}
\updated{where $\dot{\rho}_*^+$ is the star formation rate at radius
  $r$ as described in Section~\ref{sec:simulations}.}

Combining the equations in this section, we now have the expression for
the dust source and sink terms:

\begin{equation}
\ S_+ - S_- =  S_{+, inj} + S_{+, gg} -
\left(S_{-, gd} + S_{-, SF}\right) .
\end{equation}

By evolving Equation~\ref{eq:continuity} for each position and time, we can evaluate the D function in
Equation~\ref{eq:Ddef} with the computed value of $\rho_d$.

\subsubsection{Two Component Model}
\label{sec:two_component}

The one component approach assumes that dust mixes with gas
instantaneously after its creation in stellar outflows. However,
stellar ejecta contains dust not yet mixed with the ambient
ISM. Although dust-laden, these compact planetary nebulae will not
contribute significantly to the total infrared opacity of the
galaxy. Therefore, the mixing process introduces a lag time between dust
creation and its consequent effects on the radiative feedback in the
galaxy.

To estimate the mixing time, we consider a planetary nebula of mass
$\Delta M$ expanding with velocity $v_1$ relative to the parent star into a surrounding medium of
density $\rho_{\rm ext}$ and sound speed $c_s$. We further define
$v_{rel}$ as an estimate of the relative velocity of the star with
respect to the surrounding ISM, so that a fiducial value is the local
(1D) velocity dispersion of the stars. We denote
the internal density as $\rho_1$ and the radius of the nebula as
$r_1$. The nebula expands until it comes into pressure equilibrium with
the surrounding medium, i.e.,

\beq
\ v_1^2 \rho_1 = P_{ext}
~~~.
\eeq
Hence the time to reach pressure equilibrium $t_{eq}$ is

\beq
\label{eq:t1}
\ t_{eq} = \left(\frac{3 \Delta M}{4 \pi v_1 P_{ext}}\right)^{1/3}
~~~.
\eeq
Defining the Mach number ${\cal M} \equiv \frac{v_{rel}}{c_s}$, the ram pressure experienced by the expanding wind is
\beq
\label{eq:pext}
\ P_{ext} = \rho_{ext} \left(v_{rel}^2 + c_s^2\right) = v_{rel}^2 \rho_{ext} \left(1 + {\cal M}^{-2}\right)
~~~.
\eeq
Let $t_{fr}$ be the time it takes the planetary nebula to encounter a gas
mass equal to its own mass, thereby fragmenting and mixing it. Then,
\beq
\ \frac{\Delta M}{\pi r_1^2} = t_{fr} \rho_{ext}\ {\rm max}\left(v_1, v_{rel}\right)
~~~,
\eeq
where typically $v_{rel} \gg v_1$. Using Equation~\ref{eq:pext},
$t_{fr}$ can be expressed as

\beq
\label{eq:fragmentation}
\ t_{fr} = \left(\frac{16}{9\pi}\right)^{1/3} \left(1 +
  {\cal M}^{-2}\right)^{2/3} \left(\frac{v_{rel}}{v_1}\right)^{1/3}
\left(\frac{\Delta M}{v_1^3 \rho_{ext}}\right)^{1/3}
~~~.
\eeq

In addition to fragmentation, we must also consider evaporation due to
thermal conduction. Following \citet{Draine_2011} Equation~34.17,

\beq
\label{eq:conduction}
\ t_{ev} = 1.6\times10^3\ {\rm yr} \left(\frac{\Delta M}{1\
    M_\odot}\right) \left(\frac{r_{ev}}{1\ {\rm pc}}\right)^{-1}
\left(\frac{T}{10^7\ {\rm K}}\right)^{-2.5}
 ~~~,
\eeq
where $T$ is the gas temperature and $r_{ev}$ is the radius of the
nebula when it evaporates. Note that we have assumed that the Coulomb logarithm $\ln
\Lambda = 30$. In our implementation of
these equations, $\Delta M =
0.1\ M_\odot$, and $v_1 =\ 10$ km/s.

There are three distinct regimes to consider. First, consider the case
in which $t_{ev} < t_{eq}$, i.e.~the nebula evaporates before it
expands to pressure equilibrium. Then the radius $r_{ev} = v_1
t_{ev}$. Solving \updated{for $t_{ev}$}, which in
this case is the mixing time, we obtain

\begin{equation}
t_{\rm mix,1} = 8.4\times10^3\ {\rm yr}\ \left(\frac{T}{10^7\ K}\right)^{-5/4}
~~~.
\end{equation}
However, if $t_{eq} < t_{ev}$, then
$r_{ev} = r_1$. Once at pressure equilibrium, the nebula can mix via
fragmentation or evaporation, depending on which is faster. In the
former case,

\beqar
\ t_{\rm mix,2} &=& 1.5\times10^3\ {\rm yr}\ \left(\frac{T}{10^7\ K}\right)^{-5/2}
\left(\frac{v_{\rm rel}}{300\ {\rm km/s}}\right)^{2/3} \nonumber \\ &&
\times \left(1 +
  {\cal M}^{-2}\right)^{1/3}
~~~.
\eeqar
However, if $t_{\rm mix,2} > t_{fr}$, we are in
the third regime where the mixing time $t_{\rm mix,3} = t_{fr}$:

\beqar
t_{\rm mix, 3} &=& 1.9\times10^6\ {\rm yr} \left(\frac{v_{\rm rel}}{300\
    {\rm km/s}}\right)^{1/3} \left(\frac{n_{\rm H}}{0.01\ {\rm
    cm}^{-3}}\right)^{-1/3} \nonumber \\ && \times \left(1 +
{\cal M}^{-2}\right)^{2/3}
~~~.
\eeqar
In summary, the mixing time is defined as
\begin{equation}
  t_{\rm mix} = \left\{
     \begin{array}{lr}
       \displaystyle
       t_{\rm mix, 1}, &t_{ev} < t_{eq};\\
       \displaystyle
       {\rm min}\left(t_{\rm mix, 2}, t_{\rm mix, 3}\right) &t_{ev} >
       t_{eq}
       ~~~.
     \end{array}
   \right.
\end{equation}

Equipped with a mixing time, we may now modify the continuity
equations for gas and dust by distinguishing between the planetary
nebula (PN) and diffuse ISM phases. We assume that no dust growth or
destruction occurs in the PN phase.

\beqar
\label{eq:continuity_twostream}
\frac{\partial \rho_{g,ISM}}{\partial t} + \nabla \cdot \left(\rho_g
  {\bf v}\right) &=& \dot{\rho}_{\rm II} - \updated{\dot{\rho}_*^+} +
\updated{\dot{\rho}_{\rm w}} + \frac{\rho_{g,PN}}{t_{\rm
    mix}} \\
\frac{\partial \rho_{g,PN}}{\partial t} &=& \dot{\rho}_*  - \frac{\rho_{g,PN}}{t_{\rm mix}}\\
\frac{\partial \rho_{d,ISM}}{\partial t} + \nabla \cdot \left(\rho_d
  {\bf v}\right) &=& 0.0133 \dot{\rho}_{\rm II} + \nu_{gg} \rho_{d,ISM} - \nu_{gd}
\rho_{d,ISM} - \nonumber \\ 
&&\frac{\rho_{d,ISM}}{\rho_{g,ISM}} \dot{\rho}_*^+ + \frac{\rho_{d,PN}}{t_{\rm mix}}\\
\frac{\partial \rho_{d,PN}}{\partial t} &=& 0.0133 \dot{\rho}_*  -
\frac{\rho_{d,PN}}{t_{\rm mix}}
~~~,
\eeqar
where $\dot{\rho}_{\rm II}$ is the gas \updated{source term associated with
  the young stellar population via Type II supernovae and $\dot{\rho}_{\rm w}$ is the source term associated
  with winds from the circumnuclear disk. In the limit of small but
  constant $t_{\rm mix}$, we recover the gas continuity equation of
  \citet{Ciotti+Ostriker_2012}, Equation 4.73}.

\subsection{Grain Temperature}

The physics presented in the previous section directly influences the
hydrodynamical evolution of the models. Here we present additional
physics needed to compute observational properties of the models, the
other focus of this work.

By numerical integration of the radiative transfer equations in the
one-stream approximation (see Section \ref{subsec:radtrans}), we compute the
total radiation density in each shell. We approximate the total
radiation absorbed in each radial shell by
dust as the difference in luminosity at the base and end
of the shell, i.e.

\begin{equation}
\ \Delta L = L_{\rm eff, inner} - L_{\rm eff, outer} 
~~~\updated{,}
\end{equation}
\updated{where the effective luminosities include contributions from both
optical and UV bands. We assume this luminosity is radiated
by a population of grains in that shell, all with steady-state temperature
$T_d$, since the steady-state temperature of a dust grain is
nearly size-independent \citep[see e.g.][Equations 24.19 and
24.20]{Draine_2011}. I}mposing that the total
dust emission in a shell of volume $V$ is equal to the computed $\Delta L$, the relation between $T_d$ and
$\Delta L$ in a given shell is:

\begin{equation}
\label{eq:deltaL}
\ \Delta L = V \sigma
T_d^4 \int_{a_{\mathrm{min}}}^{a_{\mathrm{max}}} \! \mathrm{d}a
\frac{\mathrm{d}n_d}{\mathrm{d}a} 4 \pi a^2 Q\left(a, T_d\right) 
~~~,
\end{equation}
where $Q\left(a,T_d\right)$ is the Planck-averaged emission
efficiency. \updated{Following the power-law prescription for the
  silicate Planck-averaged emission efficiency of \citet{Draine_2011}
  Equation 24.15 at low $T_d$ and approximating the high $T_d$ behavior
  of $Q\left(a,T_d\right)/a$ as a constant, we obtain:}

\begin{equation}
\label{eq:qabs}
  Q\left(a,T_d\right) = \left\{
     \begin{array}{lr}
       \displaystyle
       \left(\frac{a}{0.1 \mu\mathrm{m}}\right) 1.3 \times 10^{-6}\ 
       T_d^2, & \, T_d < 164; \\
       \displaystyle
       \left(\frac{a}{0.1 \mu\mathrm{m}}\right) 3.5 \times 10^{-2}, &
       \, T_d >  164;
     \end{array}
   \right.
\end{equation}
where $T_d$ is in Kelvin. \updated{We note that the steady-state
  temperature for graphitic grains does not differ substantially from
  that of silicate grains \citep[][Equation 24.20]{Draine_2011},
  allowing us to focus on silicates for specificity and simplicity.}
By inserting Equation~\ref{eq:qabs} into Equation~\ref{eq:deltaL},
some algebra shows that the equilibrium dust temperature is given by

\begin{equation}
\label{eq:tdust}
  T_d = \left\{
     \begin{array}{lr}
       \displaystyle
       5.97 \left(\frac{\rho_{\mathrm{grain}} \Delta
  L}{\rho_d V}\right)^\frac{1}{6}
\hspace{0.2cm},& \hspace{0.5cm} T_d < 164; \\
\displaystyle
      1.14 \left(\frac{\rho_{\mathrm{grain}} \Delta
  L}{\rho_d V}\right)^\frac{1}{4}
\hspace{0.2cm},& \hspace{0.5cm} T_d >  164;
     \end{array}
   \right.
\end{equation}
where all quantities are in cgs. Note that the integral in Equation~\ref{eq:deltaL} is monotonic in $T_d$,
therefore ensuring only one branch of Equation~\ref{eq:tdust} is selected
for given input
values of $\Delta L$, $V$, and $\rho_d$. Because this calculation
neglects the stochastic heating of small grains, the derived grain
temperature should be considered as a characteristic temperature for
the far infrared (FIR) dust emission.

Finally, we define the luminosity-weighted dust temperature $<T_d>$ to
be
\begin{equation}
\label{eq:tdwgt}
  <T_d> = \frac{1}{L_{\rm IR}} \int  T_d\left(r\right)
  \updated{\left(\frac{\Delta L \left(r\right)}{\Delta r}\right)} {\rm d}r 
~~~,
\end{equation}
which provides an estimate of the temperature of the dust producing
the observed IR emission.

\subsection{One-Stream Radiative Transfer in Spherical Symmetry}
\label{subsec:radtrans}
 
The integration scheme for the radiative transfer equation has been
improved with respect to \citet{Ciotti+Ostriker+Proga_2010}. The full description of the new scheme is given in
\citet{Novak+Ostriker+Ciotti_2012}. In particular, we adopt the Simplified Radiation Transport
in \updated{their Appendix B}, which we describe briefly below. In
\citet{Novak+Ostriker+Ciotti_2012} the full equations of radiative
transfer were solved by using a relaxation method, and it was shown
that the following approximation works remarkably well for the present
problem.

The radiation transport equations for the black hole radiation are
particularly simple because all UV and optical photons emitted from the black hole will necessarily be
outgoing photons assuming that the scattering opacity is
negligible. This yields the relation

\begin{equation}
\ \frac{dL_{\rm eff, BH}}{dr} = - \rho \kappa_i L_{\rm eff, BH},
\end{equation}
where $\kappa_i$ is the dust opacity in band $i$ and the effective
black hole luminosity $L_{\rm eff, BH}$ is the outgoing black hole
luminosity \updated{that would be seen by an observer at radius $r$,
  i.e. after absorption.}

  \updated{We make the approximation that all absorption
  in the UV is due to dust. The photoionization opacity of the gas
  competes with the dust when the neutral fraction is above
  $\sim10^{-3}$ for Galactic dust-to-gas ratios. However, in $10^7$ K
  gas, the neutral fraction is of order $10^{-8}$ due to collisional
  ionization alone \citep[see, e.g.][Equations 14.39 and
  14.43]{Draine_2011}. During AGN ``on'' phases, a photoionizing
  luminosity of $10^{46}$ erg/s from the AGN is able to maintain a steady-state
  neutral fraction of $\sim10^{-7}$ in a dense ($n_{\rm H} = 10^3$
  cm$^{-3}$) cloud with $T = 10^4$ K at $r = 100$ pc. In both cases, the gas opacity is
  negligible compared to the dust.}

  \updated{Following \citet{Sazonov+Ostriker+Sunyaev_2004}, we assume the AGN radiates
  $10\%$ of its energy in the X-ray, $35\%$ in the UV, $25\%$ in the
  optical, and $30\%$ in the IR. The IR value includes contribution
  from a subgrid dusty accretion torus which we leave in place even in
  our ``No Dust'' model.}

For other radiation sources, however, the equation is complicated by
the fact that photons may be emitted inward toward the center of the
galaxy and, providing the optical depth is low enough, re-emerge on
the other side as an outgoing photon. To account for this, we
approximate the \updated{probability that an emitted photon will be
  outgoing, either initially or by re-emerging to the same radius on
  the other side}, by the function $\Psi$ which is defined as

\begin{equation}
\ \Psi \equiv 1 -
\frac{0.5}{1+\mathrm{exp}\left(-\tau\right)}\frac{r_1^2}{\mathrm{max}\left(r_1^2,r^2\right)}
~~~,
\end{equation}
where $\tau$ is the optical depth from $r$ to infinity and $r_1$ is the
radius at which $\tau = 1$. Using this parameterization, the radiative
transport equation for the outgoing stellar radiation $L_{\rm eff, *}$ is given by

\begin{equation}
\ \frac{dL_{\rm eff, *}}{dr} = 4\pi r^2 \Psi \dot{E}_{\updated{i}} - \rho \kappa_i L_{\rm
  eff, *} 
~~~,
\end{equation}
where $\dot{E}_{\updated{i}}$ is energy radiated by stars per unit volume per unit
time \updated{at radius $r$ in band $i$. In practice, $\dot{E}_i$ is
  dependent upon time, radius, and the local gas density and is
  partitioned into the optical and UV bands through use of
  characteristic emission efficiencies and timescales for each band
  \citep[][Equations 4.24 and 4.25]{Ciotti+Ostriker_2012}.}

\updated{The treatment of the radiation pressure has remained
  unchanged from CO07 other than the method of
  computing the dust opacity, and includes radiation pressure on gas from
  electron scattering and X-ray photoionization as well as the
  radiation pressure on dust.}

In addition to the changes detailed in
\citet{Novak+Ostriker+Ciotti_2012}, we also include an updated
prescription for the optical depth of a radial shell. If a shell is
optically thick, only a portion of the shell will experience a force
from the radiation pressure. To achieve the proper limiting behavior,
we modify the optical depth in a band $i$ of a given shell $\tau_i'$ in the following
way to obtain a $\tau_i$ for use in \updated{calculations of
  effective luminosities and radiation pressure}:

\beq
\ \tau_i \equiv 1 - {\rm e}^{-\tau_i'}
~~~.
\eeq

\section{Simulations}
\label{sec:simulations}

\updated{We present a suite of 1D hydrodynamical simulations of the
  coevolution of a giant elliptical galaxy and its central
  supermassive black hole. The simulation begins after the initial
  starburst that produced the majority of the galaxy's stellar mass,
  leaving the galaxy with no remaining gas. Cooling flow instabilities
  in the secondary gas from stellar evolution primarily drive accretion onto the central SMBH, which
leads to the production of nuclear and galactic winds. Mechanical and radiative
feedback from the AGN, Type Ia and Type II supernovae, stellar
radiation, and thermalization from stellar mass losses are all
explicitly considered.}

For simplicity, we restrict our simulations to the class of Type A
models described in \citet{Ciotti+Ostriker+Proga_2010}. In these models, the opening angle of
the broad line region (BLR) wind and the mechanical efficiency $\epsilon_w$ are independent of the
accretion luminosity. $\epsilon_w = 10^{-4}$ is a factor of two below the mechanical
efficiency assumed in many of the treatments of AGN feedback, such as
\citet{DiMatteo+Springel+Hernquist_2005}, but similar to the value
found most appropriate when winds are included \citep[see, e.g.,][]{Choi+etal_2013}. We note that including the
momentum of the outgoing wind makes a given energy input far more
effective \citep{Choi+Ostriker+Naab+Johansson_2012}. \updated{We choose to use the $A$ class for the purpose of this study as
the accretion physics is cleaner than the
more intricate $B$ class of models, whose efficiency increases with
increasing Eddington ratio \citep{Ciotti+Ostriker+Proga_2009}, and the
role of the dust is consequently easier to disentangle.}

For ease of comparison, all of the dynamical properties relevant for
the simulations is the same as in \citet{Ciotti+Ostriker+Proga_2009}, i.e. a Jaffe stellar
distribution plus a dark matter halo so that the total density
profile is proportional to $1/r^2$. The total stellar mass of $3 \times 10^{11}
M_\odot$ and effective radius $R_e = 6.9$ kpc result in a central velocity dispersion $260$ km s$^{-1}$. Dynamical properties of the model
are given in \citet{Ciotti+Morganti+deZeeuw_2009}. The
initial mass of the central SMBH is fixed to $M_{\rm BH} = 10^{-3}
M_*$ as in previous papers, therefore approximately following the
Magorrian relation. In practice, all of the evolutionary phases of
galaxy formation leading to the establishment of the Magorrian
relation are not considered. Accretion onto the BH
is computed from the full hydrodynamic equations rather than assuming
Bondi accretion or other approximate treatments. It is mediated by a
circumnuclear accretion disk whose balance equations are integrated as subgrid
physics \citep{Ciotti+Ostriker_2012}. Each simulation
employs 240 cells with the innermost gridpoint at 2.5 pc and the outermost
at 208 kpc. As in the previous papers, for simplicity we assume
standard outflow boundary conditions at the grid outer boundary and
use a dynamic time resolution based on the physical timescales in the
galaxy. However, we increase the time resolution by an additional
factor of 10 relative to previous work.

\updated{The treatment of the physics for the stellar component of the galaxy,
including stellar evolution, Type Ia and Type II Supernovae, and star
formation, as well as the hydrodynamical equations are fully described
in \citet{Ciotti+Ostriker_2012}, and we outline it briefly here. The
star formation rate at a specific radius $r$ is given by the equation}

\begin{equation}
\updated{\dot{\rho}_*^+ = \frac{\eta_{\rm form} \rho}{\tau_{\rm form}}
~~~,}
\end{equation}
\updated{where $\eta_{\rm form}$ is an efficiency coefficient dependent on the
local gas temperature and having typical values between 0.03 and 0.4
\citep{Cen+Ostriker_2006} and $\tau_{\rm form}$ is the maximum of the
gas cooling time and the dynamical time. The gas cools via Compton
cooling, bremsstrahlung, and both line and continuum cooling as
estimated by the formulae given in \citet{Sazonov+etal_2005}. These
formulae are unmodified by the inclusion of dust.}

In summary, the models are in all respect identical to
previous models with the exception of a better treatment of
dust, an improved numerical integration of the radiative
transfer \citep[see also][]{Novak+Ostriker+Ciotti_2012}, and increased
spatial and temporal resolution. However,
for the same input physics and previous dust treatment, the results
are nearly identical to previous ones. Our
$A_2$ model refers to the precise implementation of the same model in
\citet{Ciotti+Ostriker+Proga_2010} as we use Equation~\ref{eq:co_dfactor} to model
the dust depletion.

We introduce five variants of the $A_2$ model - and thus
six models in all, with each variant utilizing a
different prescription for the dust abundance and distribution. These models are
summarized in Table~\ref{table:models}, where they are listed in the
approximate order of increasing dust to gas
ratio at the end of the simulation.

In the first model, $A_2^{\rm ND}$, we consider a galaxy
completely devoid of dust, i.e. \updated{$\rho_d/\rho = 0$ at all
  radii (equivalently, the depletion factor D in Equation~\ref{eq:Ddef} is
  fixed to zero).}

$A_2^{\rm MW}$, with a dust to gas ratio equal to that of the Milky Way
scaled to the metallicity of our galaxy (Z = 4/3 Z$_{\rm MW}$),
\updated{i.e. $\rho_d/\rho = \left(4/3\right)\times10^{-2}$}, is the
other extreme model. In this maximum dust model, D = 1.

We have two additional models in which
the dust to gas ratio is a fixed number independent of time and
position. First is $A_2^{-4}$, in which $\rho_d/\rho = 10^{-4}$ at all
radii \updated{(D = $0.75\times10^{-2}$)}. This is motivated by recent
Herschel observations \citep{Smith+Gomez+Eales+etal_2011} of the dust
masses of 62 early type galaxies and scaled to our assumed stellar
mass of $3 \times 10^{11} M_\odot$. In interest of spanning the viable range of dust
to gas ratios, we also introduce $A_2^{-3}$ in which $\rho_d/\rho =
10^{-3}$ \updated{(D = $0.75\times10^{-1}$)}.

Our most sophisticated models embody the suite of physics
for grain production and destruction outlined in
Section \ref{subsec:continuity} to compute the dust
mass density at each radius and the resulting dust opacity. The
$A_2^{\rm CE}$ model employs the ``One Component''
formalism of Section ~\ref{sec:one_component} while $A_2^{\rm CE2}$ the ``Two
Component'' formalism of Section ~\ref{sec:two_component}.

\begin{table*}
\label{table:models}
\caption{Summary of Models}
    %\begin{tabular}{p{2.cm} | p{2.cm} | p{2.cm} | p{2.cm} | p{2.cm} |
      %  p{2.cm} | p{2.cm}}
\begin{center}
  %\tiny
  %\tabcolsep 5.8pt
    \begin{tabular}{|c | c || c | c | c | c | c | c | c || c | c | p{0.75cm} |
        p{0.75cm} | c | c|}
      \hline
    \multirow{2}{*}{Model} & 
    \multirow{2}{*}{Depletion}  &
    \multirow{2}{*}{$\Delta M_w$} &
    \multirow{2}{*}{$\Delta M_*$} &
    \multirow{2}{*}{$ M_{\rm gas}$} &	  
    \multirow{2}{*}{$L_X$} &
    \multirow{2}{*}{$\frac{M_{\rm dust}}{M_{\rm gas}}$} &
    \multirow{2}{*}{$\frac{L_{\rm IR}}{L_{\rm Op*}}$} &
    \multirow{2}{*}{$\tau_{\rm Op}$} &
    \multirow{2}{*}{$\Delta M_{\rm BH}$} &
    \multirow{2}{*}{$e_{\rm Bol}$} &
    \multirow{2}{*}{$e_{\rm UV}$} &
    \multirow{2}{*}{$e_{\rm IR}$} &
    \multirow{2}{*}{$N_\mathrm{burst}$} &
    \multirow{2}{*}{$f_{\mathrm{duty}}$}  \\[6pt]
    & & [M$_\odot$] & [M$_\odot$] & [M$_\odot$] & [erg/s] & &   
    & &  [M$_\odot$] & & $e_{\rm Op}$ & $e_{\rm X}$ & & \\
    \hline \hline
$\mathrm{A}_2^{\mathrm{ND}}$ & 0 & 10.43 & 9.40 & 9.04 & 38.15 & -- & -- & -- & 8.97 & 0.11 & 0.040 0.029 & 0.034 0.011 & 81 & -2.21 \\[5pt] \hline
$\mathrm{A}_2$ & 1/1+T$_4$ & 10.42 & 9.26 & 9.14 & 38.26 & \updated{-4.19} & -4.75 & -5.00 & 8.65 & 0.11 & 0.016  0.013 & 0.073 0.011 & 51 & -2.45 \\[5pt] \hline
$\mathrm{A}_2^{\mathrm{-4}}$ & $0.75\times10^{-2}$ & 10.29 & 9.38 & 9.91 & 40.49 & \updated{-4.00} & -3.68 & -3.31 & 8.64 & 0.11 & 0.023  0.020 & 0.058 0.011 & 46 & -2.25 \\[5pt] \hline
$\mathrm{A}_2^{\rm CE}$ & Computed & 10.32 & 9.36 & 9.83 & 40.47 & \updated{-3.63} & -3.15 & -2.01 & 8.67 & 0.11 & 0.013  0.010 & 0.079 0.011 & 57 & -2.42 \\[5pt] \hline
$\mathrm{A}_2^{\rm CE2}$ & Computed & 10.21 & 9.98 & 10.10 & 40.57 & \updated{-2.54} & -2.65 & -1.93 & 9.01 & 0.12 & 0.007  0.005 & 0.092 0.012 & 86 & -2.20 \\[5pt] \hline
$\mathrm{A}_2^{-3}$ & $0.75\times10^{-1}$ & 10.24 & 10.06 & 9.38 & 39.03 & \updated{-3.00} & -3.30 & -2.92 & 8.91 & 0.12 & 0.006  0.005 & 0.091 0.012 & 76 & -2.27 \\[5pt] \hline
$\mathrm{A}_2^{\mathrm{MW}}$ & 1 & 10.34 & 8.19 & 9.80 & 39.48 & \updated{-1.88} & -2.07 & -1.68 & 7.79 & 0.10 & 0.017  0.020 & 0.056 0.010 & 14 & -3.14 \\[5pt] \hline
    \end{tabular}
    \\
    \end{center}
    \textbf{Notes}: The models are arranged roughly by dust
    content, from lowest to highest. All quantities are the values
    attained at the end of the simulation, which in all cases
    represents a quiescent giant elliptical galaxy. Depletion is the ratio of dust to
    metals in the model relative to the dust to metal ratio in the
    Galaxy (Equation~\ref{eq:Ddef}); $\Delta M_{BH}$ is the
    total mass accreted by the black hole; $\Delta M_w$ is the total
    mass ejected as a galactic wind; $\Delta M_*$ is the total mass
    of new stars;
    $e_i \equiv \Delta E_i / \Delta M_{\rm BH}c^2$ is the total energy emitted by
    the black hole in band $i$ (as seen from infinity) divided by the energy equivalent of the
    black hole mass growth;
    $L_{\rm IR}/L_{\rm Op*}$ is the ratio of the IR luminosity from dust and
    the effective optical luminosity from stars; $\tau_{\rm Op}$ is
    the optical depth in the optical band; $L_X$ is the X-ray
    luminosity of the ISM in the galaxy; $N_{\rm burst}$
    is the number of burst events; and $f_{\rm duty}$ is the the fraction
    of time spent with $L_{\rm Bol} > L_{\rm Edd}/30$. All quantities
    except $e_i$ and $N_{\rm burst}$
    given as log$_{10}$.
\end{table*}

\section{A First Survey of the Models}
\label{sec:survey}

We begin by comparing the overall behavior of all models in Table~\ref{table:models}. Our
purpose is two-fold: first to understand the effects of different treatments
of the dust to gas ratio. In particular, Section \ref{subsec:burst} is
dedicated to the effects during AGN bursts. Second, we
select the subset of models that best corresponds to observations,
which we discuss in detail in Section \ref{sec:observations}. 

%%%%%%%%%%%%%%%%%%%%%%%%%%%%%%%%%%%%%%%%%%%%%%%%%%%%%%%%%%%%%%%%%%%%%%%%%%%
% Figure-- Black Hole Mass and Wind Mass
\begin{figure}[htp]
 \centering
  \scalebox{0.65}{\includegraphics[width=14.0cm,angle=0,
                 clip=true,trim=0.2cm 1.0cm 0.5cm 0.0cm]{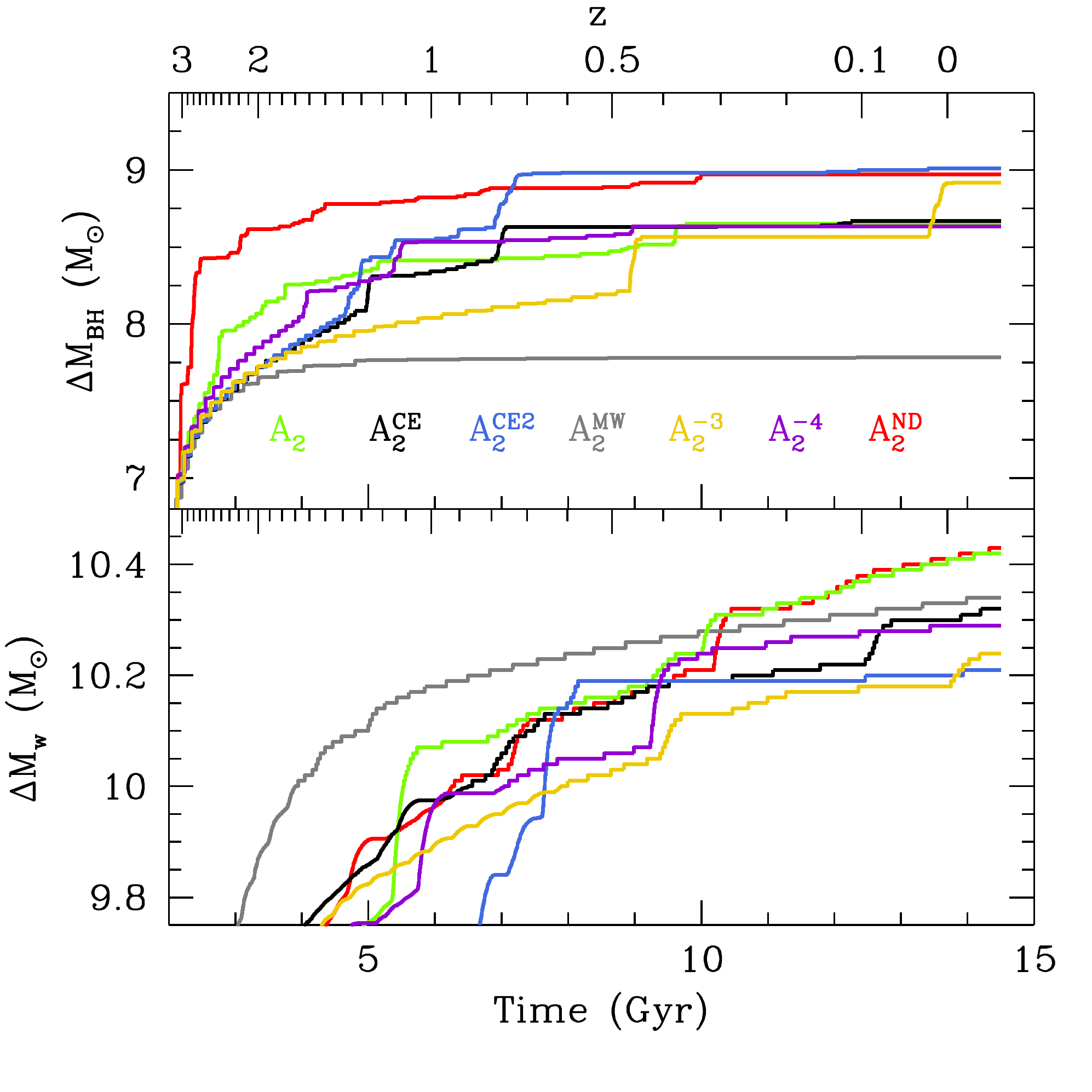}}
\caption{{\em Top}: The black hole mass growth since the beginning of
  the simulation.  {\em Bottom}: The total mass ejected as a galactic wind. AGN activity peaks at early times (z
  $\sim$ 2-3) in all models, and the black holes are quiescent in all
  models by the present epoch (z $\sim$ 0). All y-axis quantities are given as
    $\log_{10}$.} \label{fig:mbh} 
\end{figure}
%%%%%%%%%%%%%%%%%%%%%%%%%%%%%%%%%%%%%%%%%%%%%%%%%%%%%%%%%%%%%%%%%%%%%%%%%%%

%%%%%%%%%%%%%%%%%%%%%%%%%%%%%%%%%%%%%%%%%%%%%%%%%%%%%%%%%%%%%%%%%%%%%%%%%%%
% Figure-- Dust Mass
\begin{figure}[htp]
 \centering
  \scalebox{0.65}{\includegraphics[width=14.0cm,angle=0,
                 clip=true,trim=1.5cm 2.0cm 1.0cm 0.0cm]{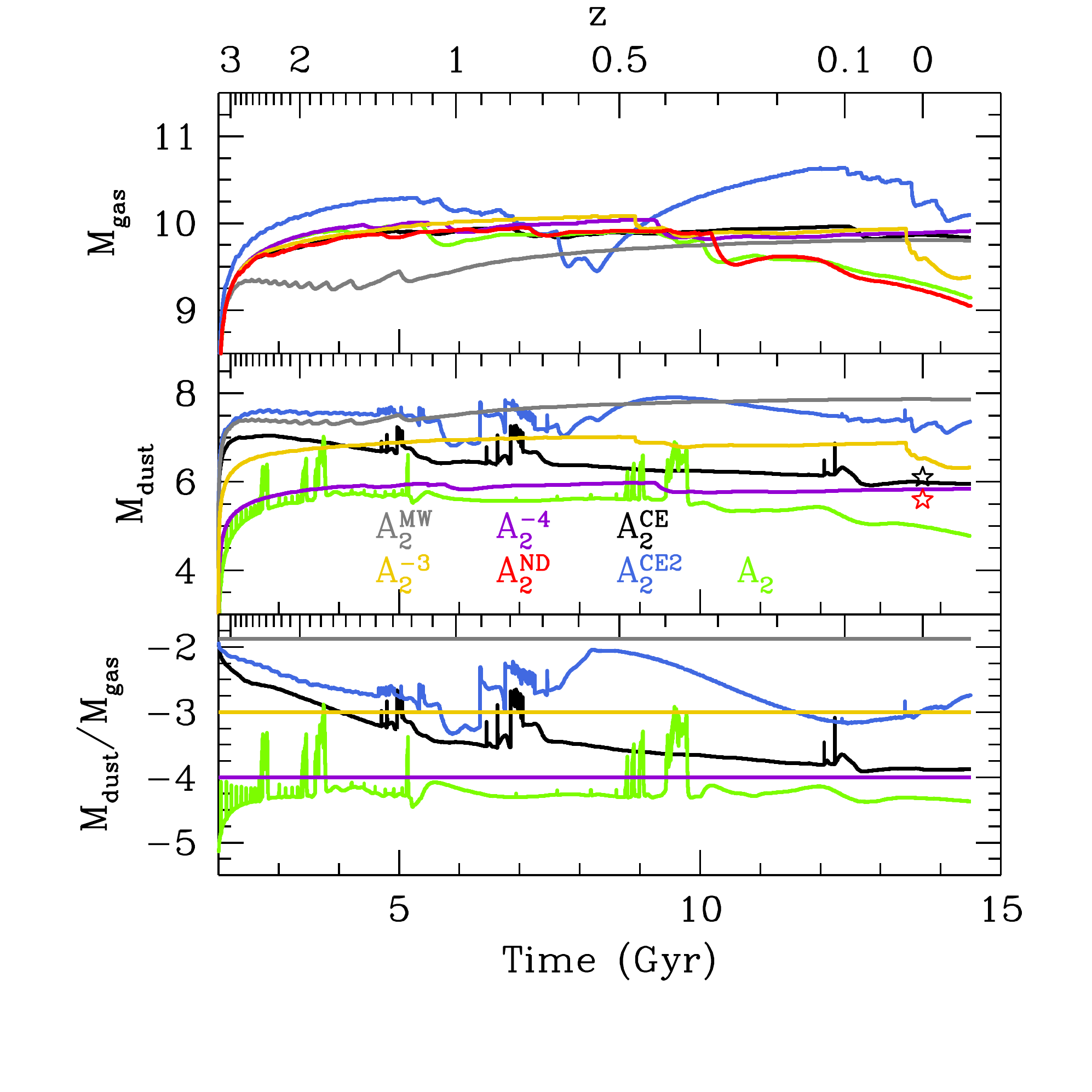}}
\caption{{\em Top}: The total gas content of the galaxy in $M_\odot$. The colors
  are the same as in Figure \ref{fig:mbh}. {\em Middle}: The total
  dust content of the galaxy in $M_\odot$. Spikes occur during cooling
  instabilities, leading to the formation of infalling shells prior to
  outbursts. We plot the average dust mass for early type galaxies as
  determined by the Herschel Reference Survey \citep{Smith+Gomez+Eales+etal_2011} as a
  black star (detections only) and red star (including
  non-detections). {\em Bottom}: The dust to gas ratio of the galaxy. All y-axis quantities are given as
    $\log_{10}$.} \label{fig:dustmass} 
\end{figure}
%%%%%%%%%%%%%%%%%%%%%%%%%%%%%%%%%%%%%%%%%%%%%%%%%%%%%%%%%%%%%%%%%%%%%%%%%%%

The first column of Table~\ref{table:models} shows the mass ejected as
a galactic wind, illustrating that
the bulk of the mass produced by stellar evolution is ejected as a
galactic wind (see also Figure~\ref{fig:mbh}). Such galactic winds in our model are supported by
thermalization of stellar motion and in particular by heating provided
by Type Ia supernovae. Due to the time dependence of the supernovae
and star formation, the specific heating rate increases with
time.

\updated{Earlier work by \citet{Renzini+Ciotti_1993} and
  \citet{Ciotti+Ostriker_2001} has shown that Type Ia supernovae are
  capable of driving winds from the outer parts of elliptical galaxies
  but have little effect on the inner $\sim 1$ kpc region. Within this
  radius, feedback from the central AGN prevents continual infall and
  can drive material to radii where supernova winds dominate.}

\updated{The correlations we observe with dust abundance are tied to
  this assisting role of the black hole. AGN feedback is more
  effective when there is more dust due to increased radiation
  pressure, and thus models with high dust content, such as $A_2^{\rm
    MW}$, are able to drive out more mass in winds early in the
  simulations during periods of intense bursting. However, the dust
  abundance can have the opposite effect at late times since the black
  hole is not able to accrete dusty gas as effectively and bursting
  may stop. Consequently,
  the $A_2^{\rm MW}$ model has relatively little wind at late times
  whereas models with little dust continue ejecting mass throughout
  the duration of the simulation. Additionally, larger black holes
  have higher Eddington luminosities and are thus more effective in
  driving winds, and the black holes grow more in models with little
  dust. Thus, dust-rich models tend to eject mass in winds at early
  times more so than dust-poor models, whereas dust-poor models have
  significantly more winds at late times.}

The next column reports the mass of new stars formed over the
simulation. Note that this value is always intermediate between
$\Delta M_{\rm BH}$ and $\Delta M_w$. This fact has important
cosmological implications as it clearly shows how Type Ia supernovae
are responsible for the metal pollution of the IGM since the bulk of
the gas is ejected, not locked into new stars. As already described in
CO07, AGN feedback has competing effects on star formation, acting as
both positive feedback during bursts and as negative feedback at the
end of each burst. This leads to the surprising result that pure
cooling flow models may form fewer stars than models with AGN
feedback. It is also known that, at least in 1D models, the bulk of
star formation happens in a region of about a few hundred parsecs in
size where cold shells are formed by recurrent cooling instabilities
and shocks induced by AGN feedback. Therefore, we expect a correlation
between the number of bursts and the number of new stars formed (see
Table~\ref{table:models}).

In the next column, we report the total amount of gas in the
simulation. Overall, the gas masses are consistent with observations
of galaxies with comparable velocity dispersions
\citep{Canizares+Fabbiano+Trinchieri_1987,
  Kim+Pellegrini_2012}. \updated{The correlation between gas mass and
  dust abundance is tied to both the ability of dust to prevent accretion
  and the more nuanced effects of dust on galactic winds discussed
  above. The $A_2$ and $A_2^{\rm ND}$ models, which have settled into
  an outflow state by the end of the simulation, have the least gas.}

In the next column is the final X-ray luminosity of the hot gaseous
corona of our models obtained by integrating the gas emissivity in the
0.38 keV band within the volume of 10 effective optical radii. For all
models, luminosities are in the observed range
\citep{Boroson+Kim+Fabbiano_2011}. The low luminosities of models $A_2^{\rm ND}$ and $A_2$, in
conjunction with their low total gas mass,
demonstrate that these models are in a global wind phase at the
present time. The
luminosities of the other models are consistent with
inflow/partial wind states. At the end of the simulation, all models
are in a state of hot, low-luminosity accretion.

In the next column, we give the dust to gas ratio over the
galaxy at the end of the simulation. The models are ordered as
expected from the dust physics. By construction, $A_2^{\rm ND}$ is
inconsistent with observations since it has no dust, and $A_2^{\rm MW}$
has too much dust relative to observed giant
ellipticals. Additionally, $A_2^{-3}$ and $A_2^{\rm CE2}$ are on the
high end of what would be expected (see
Section \ref{subsec:etgdust}). However, the enhanced dust content of the
$A_2^{\rm CE2}$ model is not directly related to the dust treatment,
but rather is due to a large star formation episode following the last
AGN bust (see bottom panel of Figure~\ref{fig:dustmass}). More
extensive exploration of the $A_2^{\rm CE2}$ model is needed to
determine if these star formation episodes are a generic feature of
the model.

The next column reports the ratio of the IR dust emission from
reprocessed optical and UV radiation from stars and the black hole to
the total effective luminosity of stars in the optical band. This ratio varies widely between
models, and is thus an important observational diagnostic. The ratio
has the expected behavior-- as the dust abundance increases, the IR
luminosity increases and the optical luminosity decreases due to
absorption. Thus, the ratio should increase with increasing dust,
which is the observed behavior. This trend is also evident in the next
column, which gives the optical depth in the optical band to the
center of the galaxy. For
elliptical galaxies, \citet{Smith+Gomez+Eales+etal_2011} find a ratio
of FIR to B-band luminosity $-2.5 < \log{L_{\rm FIR}/L_{\rm B}} <
-1.5$ with a number of upper limits at the lower end of the range. We
stress that our $L_{\rm IR}/L_{\rm Op*}$ does not correspond exactly,
but we can still make some useful comparisons. If we include upper
limits, all models are in agreement. However, when restricting the
comparison to detections, $A_2^{\rm ND}$ and $A_2$ are clearly ruled
out and $A_2^{-4}$ and $A_2^{\rm CE}$ are only marginally consistent.

In summary, all models produce acceptable results from a hydrodynamic
point of view. We can however exclude models based on their dust
content-- $A_2^{\rm ND}$ and $A_2^{\rm MW}$ clearly have too little
and too much dust, respectively, and there is tension between
observations of dust in elliptical galaxies the high
dust content of models $A_2^{-3}$ and $A_2^{\rm CE2}$.

Now we discuss the energetic aspects of black hole accretion. The next
column illustrates a factor of $\simeq 10$ spread in black hole
growth by the end of the simulations. The mass growth is strongly
correlated with the dust to gas ratio of the galaxy, with low dust
models having more black hole growth. Dust grains, which have
large UV absorption cross-sections, absorb UV photons and thus momentum
from the luminous black hole. This radiative momentum in turn props up the
gas, retarding its rate of accretion. Thus, the presence of dust tends to
screen the black hole from accreting gas. Similarly, radiative
feedback is able to more effectively terminate bursting events in
models with more dust.

A simple check on the validity of a given model is whether the final
black hole mass is consistent with the $M_{\rm BH}-\sigma$
relation or whether the final black hole mass is too large. Stellar
evolution over a cosmological time releases an amount of gas into the
galaxy equal to $\simeq30\%$ of the initial stellar mass. If more than
$\simeq1\%$ of this gas were to be accreted, the $M_{\rm BH}-\sigma$
relation would be violated. The black hole growth in our models never
exceeds $\simeq10^9\ M_\odot$ (see
Fig.~\ref{fig:mbh}, top panel), which preserves the Magorrian relation
we assumed at the outset of the simulation

The next three columns of Table~\ref{table:models} give the integrated
effective luminosity in the indicated band in units of $\Delta M_{\rm
  BH} c^2$. We recall that the adopted the electromagnetic efficiency
of our simulations is ADAF-like, declining at low accretion rates and
saturating to a prescribed value at high accretion rates. We use a
saturation value of 0.125 (see CO07 Equation~33). Since the values of
$e_{\rm Bol}$ are roughly constant and near the saturation value, the
bulk of accretion must occur at high accretion rates independent of
the dust treatment. However, the distribution into different bands is
sensitive to dust due to opacity effects.

$e_{\rm UV}$ indicates the amount of dust during periods of high quasar
luminosity, so it is naturally maximal in the no dust model $A_2^{\rm ND}$. The optical output
tends to follow the UV in its overall behavior. The hard X-ray output
is very similar in all models since the dust plays no part in its
transmission. However, this component is slightly lower in the
$A_2^{\rm MW}$ model 
since the maximal dust model emits a larger fraction of its energy at
low Eddington ratios where the overall radiative efficiency is lower
in the A type of models.

Two of the energy output columns allow us to discriminate cleanly
among the models, eliminating those having observational properties
inconsistent with known data. One important ratio is that of
the total quasar electromagnetic output to the observed AGN optical, as
inferred by \updated{$e_{\rm Bol}/e_{\rm Op}$}. This
ratio, the ``bolometric correction'', has been classically estimated
to be in the range of 5 to 10 \citep{Soltan_1982,
  Yu+Tremaine_2002}. \citet{Richards+Lacy+Storrie-Lombardi+etal_2006}
created composite SEDs of 249 quasars using photometry from Spitzer
and the Sloan Digital Sky Survey. They measured a mean ratio of
bolometric luminosity to total
optical luminosity (integrated from 0.1 to 1 $\mu$m) of $2.9 \pm 1.5$,
with values ranging between 1.8 and 19, and a bolometric correction to the
$5100 \AA$ flux of $10.3 \pm 2.1$. Most of our models fall comfortably
within the 5 -10 range, though the $A_2^{\rm CE2}$ and $A_2^{-3}$
models have a bolometric correction exceeding 20.

Additionally, \citet{Richards+Lacy+Storrie-Lombardi+etal_2006} report
an integrated IR flux between 1 and $100 \mu$m for their quasar
sample, with no corrections made for the ISM of the host galaxy. The
ratio of the mean
bolometric luminosity to the integrated IR luminosity is $2.58 \pm
0.75$, and IR to optical ratio of $1.3 \pm 1.5$. These ratios
spanned a range of 1.1 to 5.8 and 0.36 to 18, respectively. With the
exception of the $A_2^{\rm ND}$ model, all models have total IR
($e_{\rm IR}$)
exceeding total optical ($e_{\rm Op}$) by a factor greater than two and as much as
8. $A_2^{\rm CE2}$, $A_2^{-4}$, and $A_2^{\rm MW}$ have values
closer to the mean. For the radiation output from the AGN itself, we have
implicitly assumed a total IR to optical ratio of 1.2. Deviations from
this value are due entirely to processing by the galaxy.

\begin{figure*}[htp]
 \centering
 \begin{tabular}{cc}
  		\scalebox{0.40}{\includegraphics{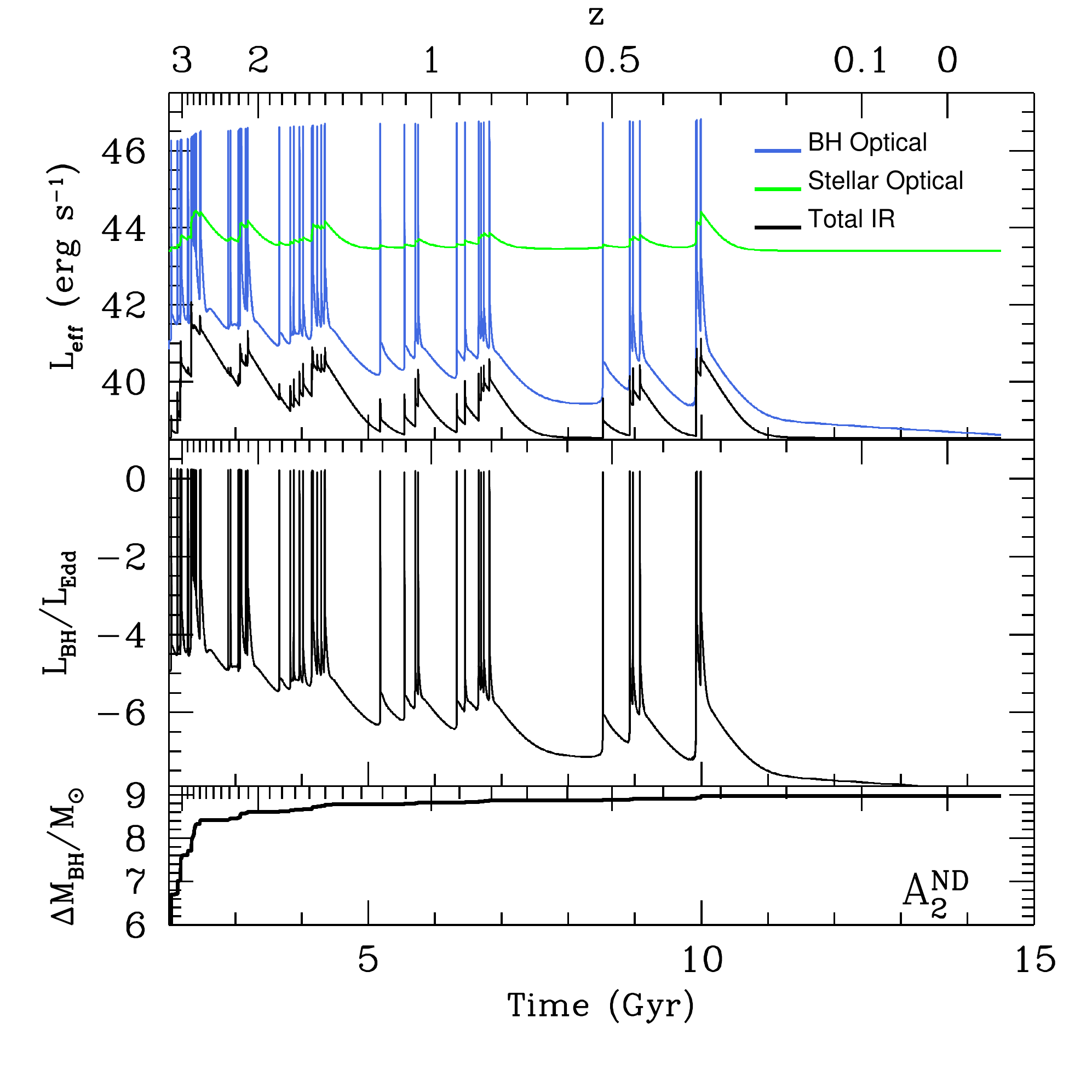}} &
 		\scalebox{0.40}{\includegraphics{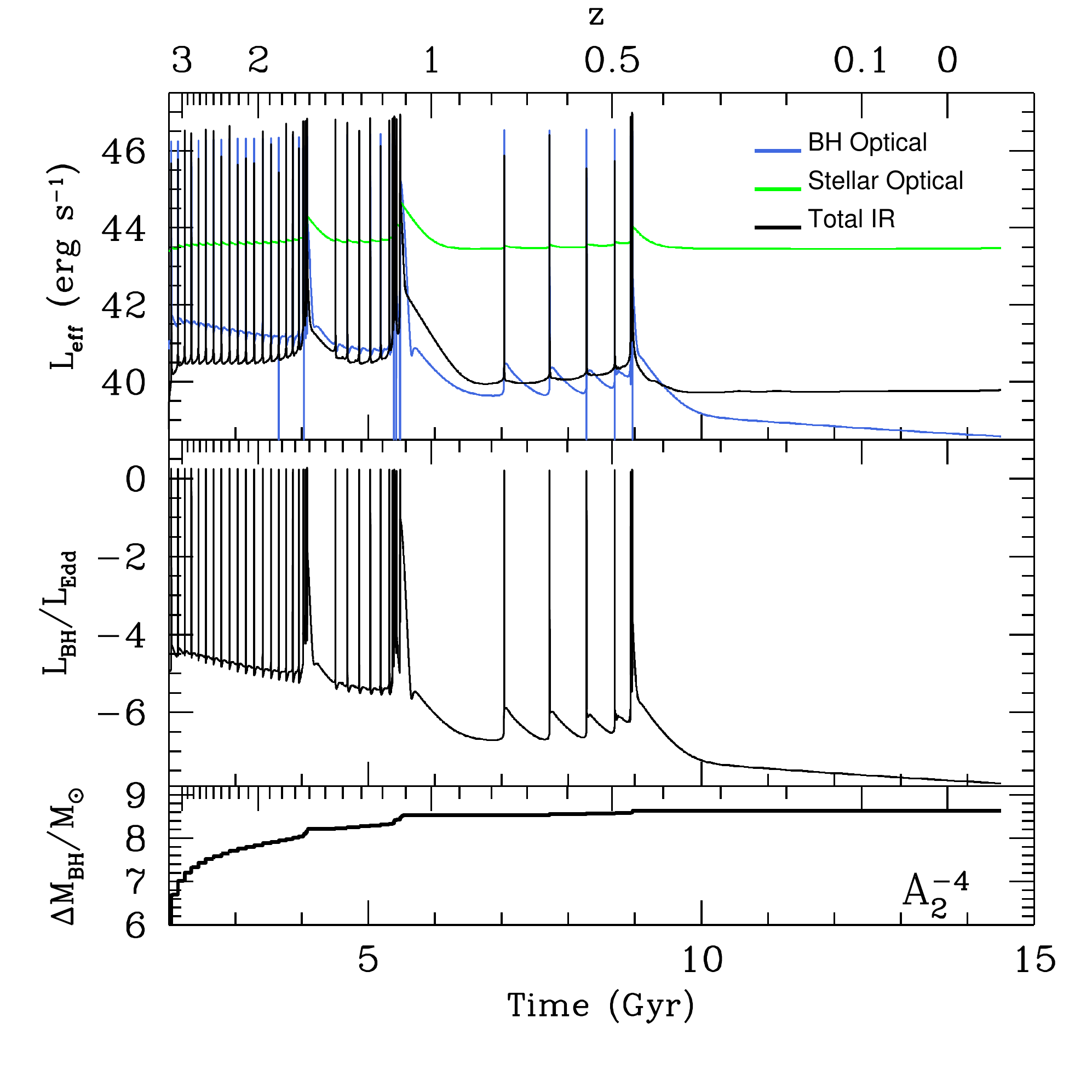}} \\
  		\scalebox{0.40}{\includegraphics{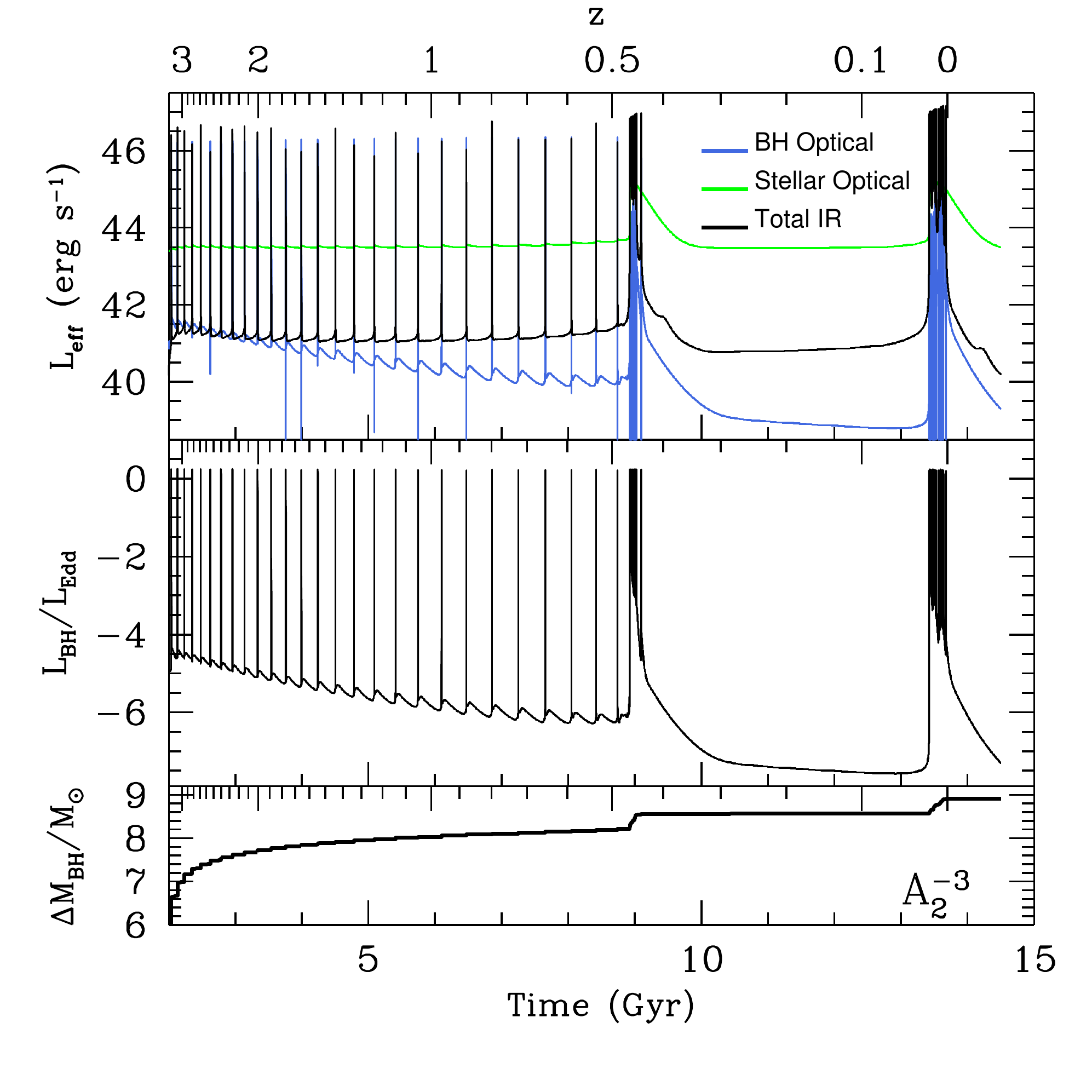}} &
 		\scalebox{0.40}{\includegraphics{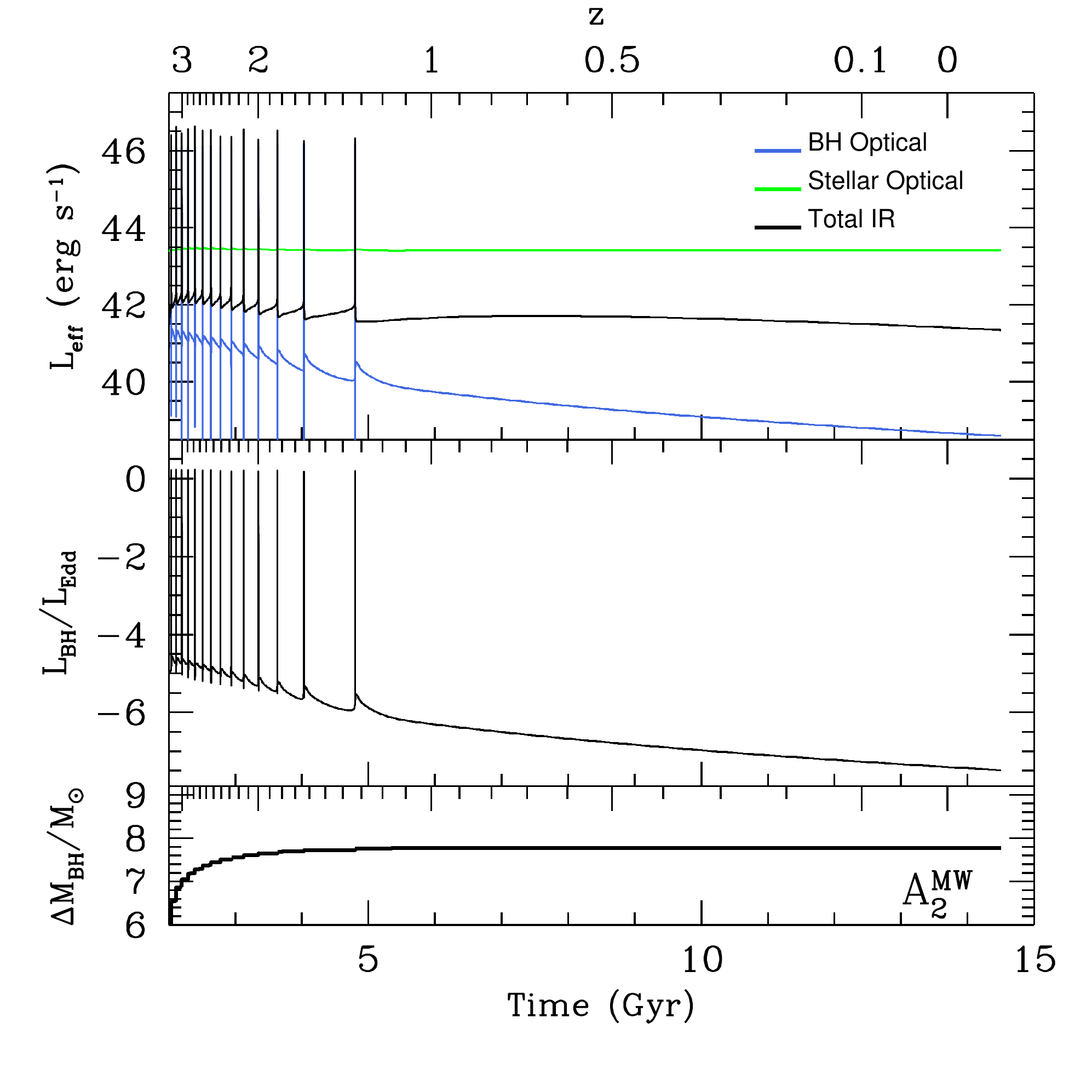}}
              \end{tabular}
\caption{A comparison of the luminosity evolution for all models
  with constant dust to gas ratios. {\em Clockwise
    from top left}: $A_2^{\rm ND}$, $A_2^{-4}$,
  $A_2^{\rm MW}$, and $A_2^{-3}$. Each figure is organized as
  follows. \emph{Top}: the luminosity seen at infinity in the optical
 from the black hole (blue) and the stars
 (green). The total IR luminosity, including the contribution from the
 central black hole,
 is plotted in black. \emph{Middle}: The Eddington fraction, defined as
 the bolometric black hole luminosity divided by the Eddington
 luminosity. \emph{Bottom}: The total black hole growth since the
 beginning of the simulation. All quantities are given as log$_{10}$.}  \label{fig:lumconst} 
\end{figure*}

\begin{figure}[htp]
 \centering
  \scalebox{0.5}{\includegraphics[width=14.0cm,angle=0,
                 clip=true,trim=0.5cm 1.0cm 0.5cm 0.0cm]{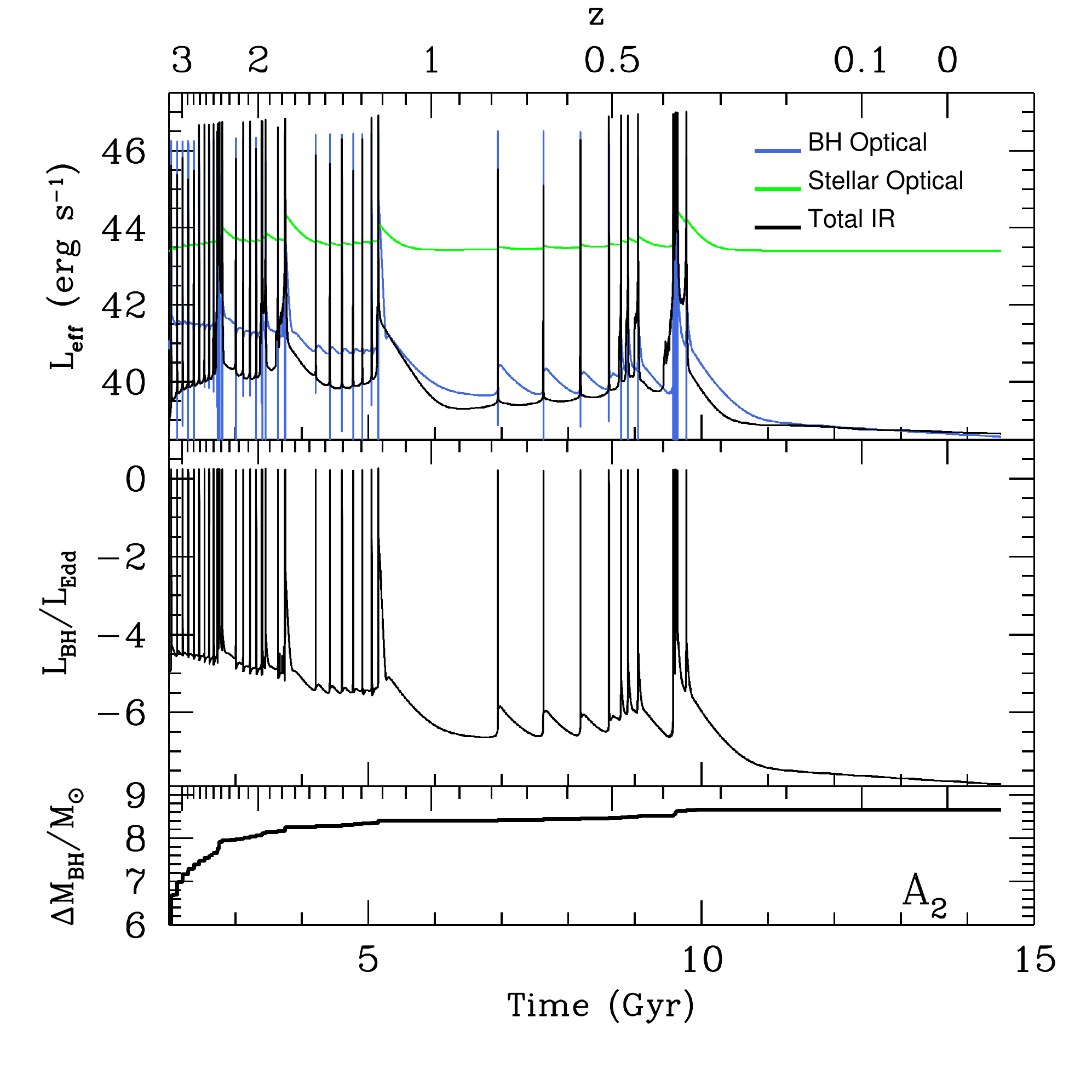}}\\
  \scalebox{0.5}{\includegraphics[width=14.0cm,angle=0,
                 clip=true,trim=0.5cm 1.0cm 0.5cm 0.0cm]{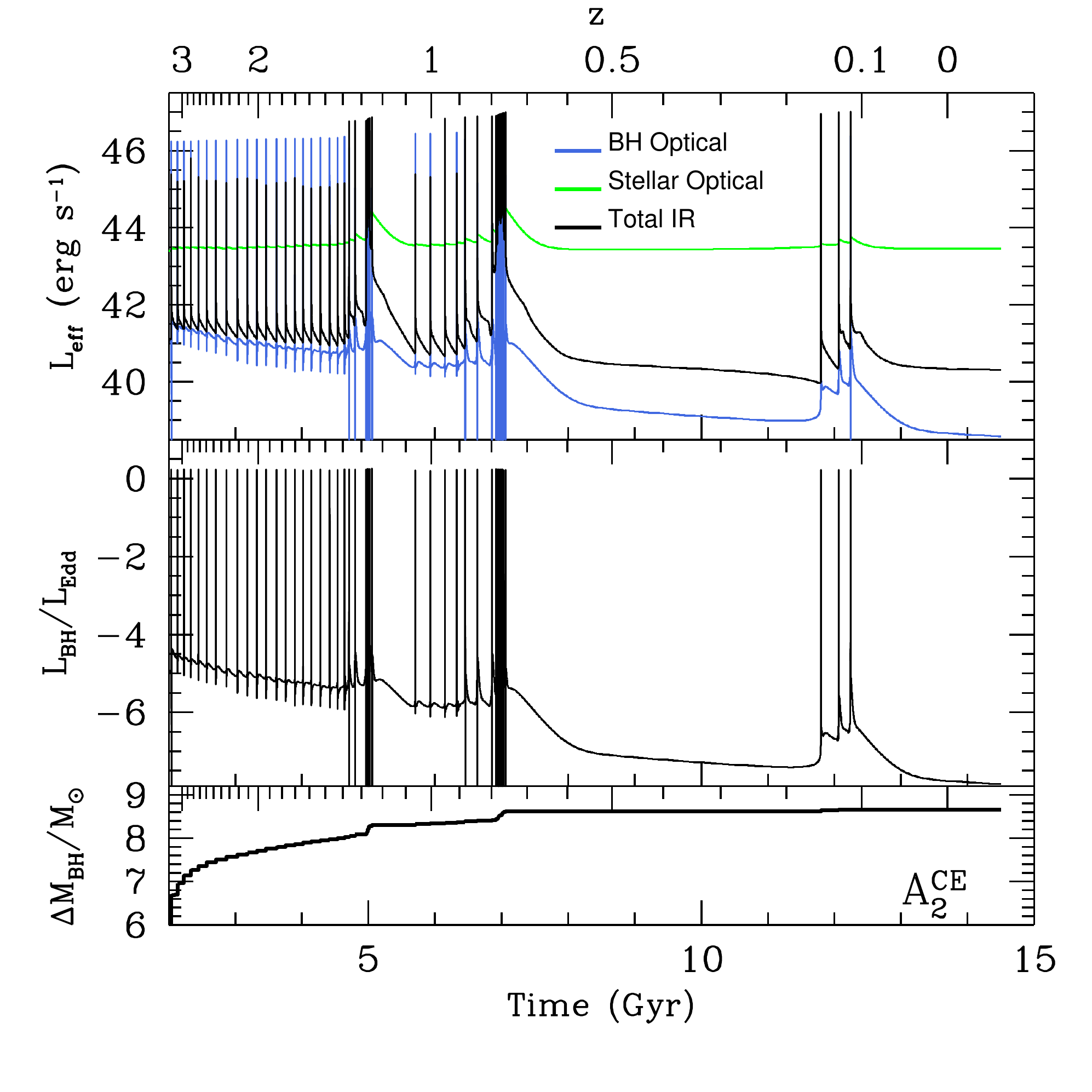}}\\
  \scalebox{0.5}{\includegraphics[width=14.0cm,angle=0,
                 clip=true,trim=0.5cm 1.0cm 0.5cm 0.0cm]{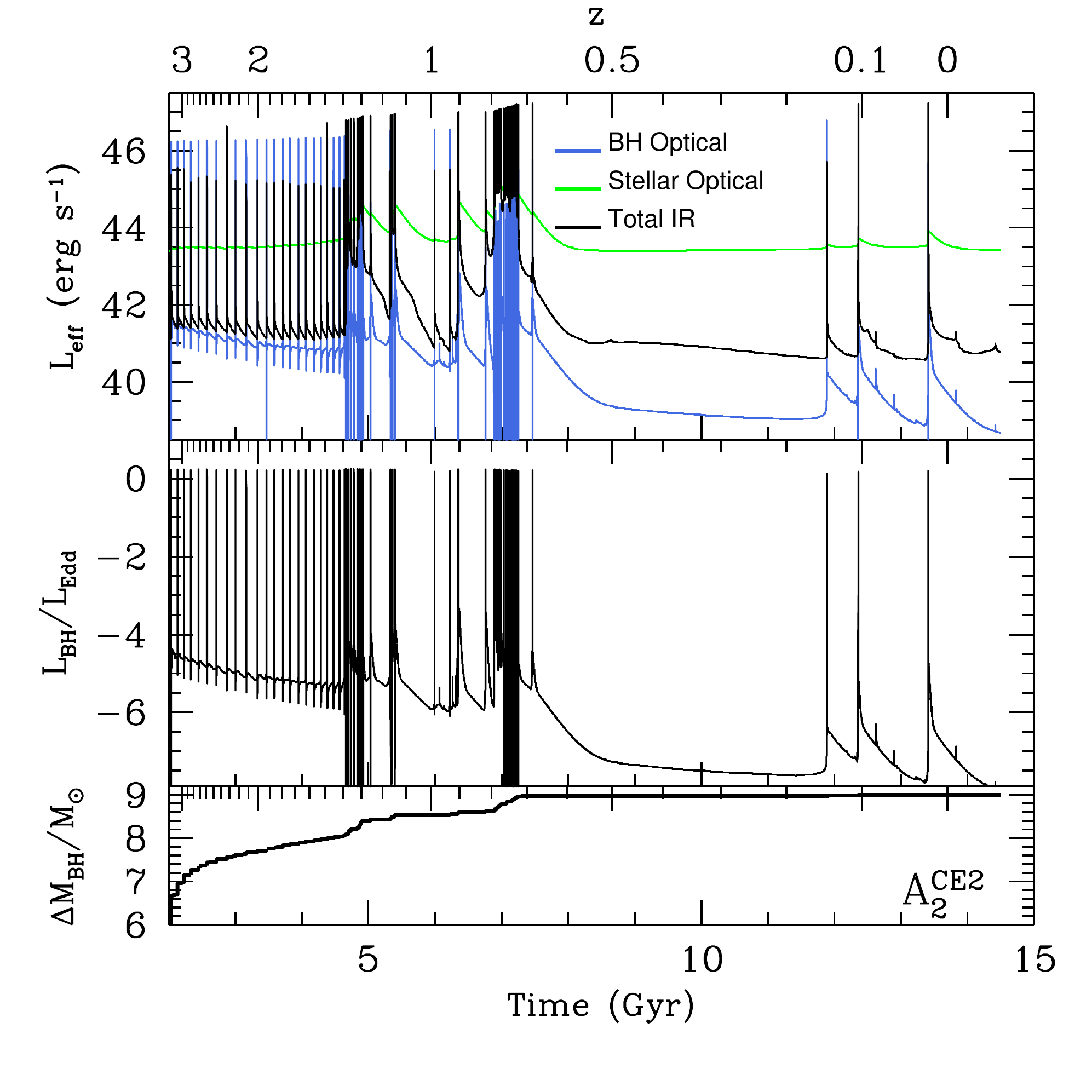}}
\caption{As in Figure~\ref{fig:lumconst}, but for the $A_2$ and continuity models
  which solve for the dust abundance as a function of radius. All three
  models have sharp continual bursts throughout the simulation as well
as significant black hole growth, consistent with the models with low
dust abundance.} \label{fig:lumvar} 
\end{figure}

In Figures~\ref{fig:lumconst} and \ref{fig:lumvar}, we present the evolution of
the effective optical luminosities of the black hole and stars, the
total IR luminosity ($L_{\rm IR}$ from Table~\ref{table:models} plus
a contribution from the central black hole), the Eddington fraction,
and the black hole mass. The two figures consider separately models
with constant dust to gas ratios and those where this ratio varies
with time and radius. The top panel of Figure~\ref{fig:lumvar} shows the
evolution of the $A_2$ model taken from \citet{Ciotti+Ostriker+Proga_2009} with the improvements
detailed in Section \ref{sec:simulations}. The time evolution of each simulation has some variation from model to
model, but the AGN activity of all models declines with cosmic
time. This decline demonstrates how the main driver of secular
evolution is the relative importance of mass injection (declining as
$\approx t^{-1.4}$) and supernova heating (declining as $\approx
t^{-1}$), so that the specific heating of the galaxy declines and
galaxies develop a global wind. The sharpness of the bursts is due to
the use of the A family of models, which have sharper bursts and
shorter duty cycles than the B family. All differences above these general
trends are due to the treatment of dust.

%%%%%%%%%%%%%%%%%%%%%%%%%%%%%%%%%%%%%%%%%%%%%%%%%%%%%%%%%%%%%%%%%%%%%%%%%%%
% Figure-- Constant D/G: Luminosity Evolution
%%%%%%%%%%%%%%%%%%%%%%%%%%%%%%%%%%%%%%%%%%%%%%%%%%%%%%%%%%%%%%%%%%%%%%%%%%%

%%%%%%%%%%%%%%%%%%%%%%%%%%%%%%%%%%%%%%%%%%%%%%%%%%%%%%%%%%%%%%%%%%%%%%%%%%%
% Figure--Variable D/G: Luminosity Evolution
%%%%%%%%%%%%%%%%%%%%%%%%%%%%%%%%%%%%%%%%%%%%%%%%%%%%%%%%%%%%%%%%%%%%%%%%%%%

The duty cycles shown in the last column of Table~\ref{table:models} are somewhat shorter than the 0.01
typical of current observations. As previously discussed, 1D
simulations have inherently less steady accretion due to the inability
of the gas to fragment. Additionally, the $A$ class of models has
routinely produced short duty cycles due to its fixed efficiency for driving
winds resulting in short duration bursts. In contrast, the
more intricate $B$ class of models typically produced higher duty
cycles \citep{Ciotti+Ostriker+Proga_2009}. 

\begin{figure}[htp]
 \centering
  \scalebox{0.65}{\includegraphics[width=14.0cm,angle=0,
                 clip=true,trim=0.5cm 1.0cm 0.5cm 0.0cm]{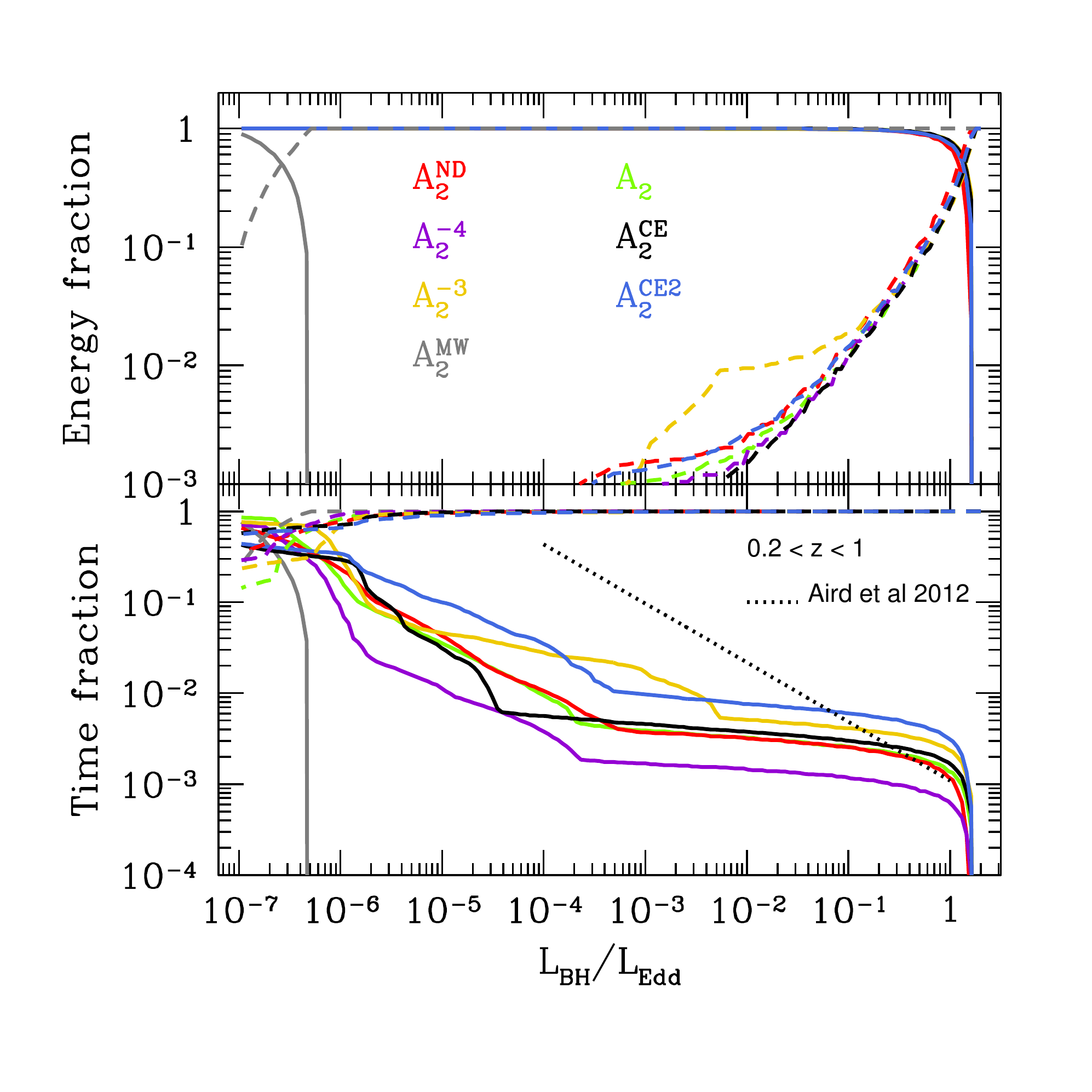}}
\caption{\emph{Top}: The fraction of energy emitted above a given fraction of
  Eddington luminosity in the time interval $0.2 < z < 1$. \emph{Bottom}: The fraction of time spent above a given fraction of
  Eddington luminosity. In both plots, we consider the bolometric
  black hole luminosity. For comparison, we plot the best-fit
  model of \citet{Aird+etal_2012} for the same time window.} \label{fig:eddfrac} 
\end{figure}

Figure~\ref{fig:eddfrac} shows that in all models, most of the energy
is emitted at or above the Eddington limit. However, the amount of time spent at a
given fraction of Eddington varies substantially among the dust models
considered here, with very low
duty cycles being typical. A very small fraction of the time in all models
is spent above $L_{\rm Edd}$. To gauge how
much time each model spends in a quiescent phase, we also plot the
amount of time spent below a given fraction of
Eddington. \citet{Ho_2009} finds that roughly 50\% of AGN have
$L_{\rm BH}/L_{\rm Edd} < 10^{-5}$. As the dust content of the models goes down, the time
spent at high Eddington fraction increases. This supports the idea
that gas is more easily able to stream to the center of the galaxy in
low dust models, resulting in sharp luminous bursts. In contrast, high
dust models require more gradual buildup of cold dense shells of
infalling gas before being able to overcome the radiative pressure
exerted by the central black hole.

\citet{Aird+etal_2012} find that the probability density function of
finding a galaxy with a specific Eddington ratio is well-described by
a power law between Eddington fractions of $10^{-4}$ and 1. We plot
the corresponding cumulative distribution function for comparison in
Figure~\ref{fig:eddfrac} and find again that our 1D models spend too
little time at Eddington ratios of $10^{-4} - 10^{-1}$. However, the
data is in rough agreement with the simulations for very high
Eddington ratios.

%%%%%%%%%%%%%%%%%%%%%%%%%%%%%%%%%%%%%%%%%%%%%%%%%%%%%%%%%%%%%%%%%%%%%%%%%%%
% Figure-- Eddington Fraction
%%%%%%%%%%%%%%%%%%%%%%%%%%%%%%%%%%%%%%%%%%%%%%%%%%%%%%%%%%%%%%%%%%%%%%%%%%%

\begin{table*}
\label{table:lir}
\caption{IR Duty Cycle}
\begin{center}
    \begin{tabular}{|c || c | c || c | c || c | c || c | c || c | c || c
      | c || c | c}
      \hline
    Model & \multicolumn{2}{|c|}{z: 3 - 2.5} &
    \multicolumn{2}{|c|}{z: 2.5
      - 2} & \multicolumn{2}{|c|}{z: 2 - 1.5} &
    \multicolumn{2}{|c|}{z: 1.5
      - 1} & \multicolumn{2}{|c|}{z: 1 - 0.5} &
    \multicolumn{2}{|c|}{z: 0.5
      - 0} \\
    \hline \hline
${\rm A}_2^{\rm ND}$ & 0.00 & 0.00 & 0.00 & 0.00 & 0.00 & 0.00 & 0.00 & 0.00 & 0.00 & 0.00 & 0.00 & 0.00\\[5pt] \hline
${\rm A}_2$ & 0.23 & 0.08 & 1.14 & 0.71 & 0.76 & 0.49 & 0.16 & 0.09 & 0.03 & 0.01 & 0.29 & 0.22\\[5pt] \hline
${\rm A}_2^{-4}$ & 0.31 & 0.10 & 0.22 & 0.07 & 0.75 & 0.46 & 1.33 & 0.47 & 0.04 & 0.01 & 0.07 & 0.05\\[5pt] \hline
${\rm A}_2^{\rm CE}$  & 0.44 & 0.02 & 0.39 & 0.00 & 0.34 & 0.00 & 0.89 & 0.58 & 0.75 & 0.52 & 0.04 & 0.02\\[5pt] \hline
${\rm A}_2^{\rm CE2}$ & 0.44 & 0.00 & 0.39 & 0.01 & 0.39 & 0.00 & 2.03 & 1.20 & 11.60 & 1.27 & 0.06 & 0.03\\[5pt] \hline
${\rm A}_2^{-3}$ & 0.46 & 0.13 & 0.36 & 0.12 & 0.22 & 0.06 & 0.18 & 0.06 & 0.15 & 0.05 & 5.19 & 0.91\\[5pt] \hline
${\rm A}_2^{\rm MW}$ & 0.50 & 0.36 & 0.34 & 0.22 & 0.11 & 0.07 & 0.04 & 0.02 & 0.00 & 0.00 & 0.00 & 0.00\\[5pt] \hline
    \end{tabular}
    \\
    \end{center}
    \textbf{Notes}: Each cell contains the percentage of time during a
    given redshift range that the galaxy has an IR luminosity
    comparable to LIRGs ($10^{11} L_\odot$, left) and ULIRGs ($10^{12}
    L_\odot$, right). In most models, phases of intense IR output have
    petered out below redshift $\simeq1$ and in nearly all by $\simeq0.5$.
\end{table*}

Taken together, these aspects of the models clearly identify those
that are unphysical. The $A_2^{\rm ND}$ model produces too little IR emission
while the $A_2^{\rm MW}$ produces too much. The standard $A_2$
model at late times has very little IR output
relative to optical, suggesting that it \updated{has too little dust}
given its \updated{stellar} mass. $A_2^{-4}$, \updated{which has} similar levels of
depletion \updated{relative to the Galactic dust to metals ratio (D
  $= 0.75\times10^{-2}$)}, has an order of magnitude greater IR output as it also has
roughly an order of magnitude more gas and therefore dust, bringing
its value of $L_{\rm IR}/L_{\rm Op*}$ closer to the observational
value. However, the $A_2^{-4}$ model does not have adequate dust to
produce significant obscuration, which is inconsistent with high
obscured fractions. Like the $A_2^{-4}$ model, both models employing the
continuity equation treatment of the dust abundance have a reasonable
amount of IR emission, in addition to having sensible values for both
the black hole mass and the duty cycle. In conclusion (and perhaps not
surprisingly), the $A_2^{\rm CE}$ and $A_2^{\rm CE2}$ models pass the
preliminary screenings better than the other models, and we will focus
on these models in Section \ref{sec:observations}.

\subsection{Burst Behavior}
\label{subsec:burst}

Due to the relevance of the black hole accretion physics, we now
discuss the burst behavior. Each burst begins with the formation of a
cooling gas shell at $\lesssim1$ kpc from the center of the galaxy. In 1D simulations, this cold
shell starts to fall toward the center and compresses the gas interior
to it. As the gas density is increased, the black hole luminosity also
increases. As soon as the black hole reaches $\simeq0.01 L_{\rm Edd}$,
pre-heating instabilities appear and the accretion becomes unstable
with shock waves propagating toward the falling cold shell. Fresh
material is carried to the black hole by reflected shock waves. The
gas in the cold shell is compressed and star formation is
induced. However, the piling up of cooling material from outside the
shell pushes the cold material to the center. The accretion of this
material produces a large final accretion event that quenches star
formation.

This general evolution is naturally affected by gas opacity, which
determines how well radiation pressure works against the falling
shell. Therefore, it is not surprising that the details of each burst
change with the different dust treatments.

\begin{figure*}[htp]
  \centering
  \scalebox{0.6}{\includegraphics[width=14.0cm,angle=0,
                 clip=true,trim=1.0cm 1.0cm 8.8cm 0.0cm]{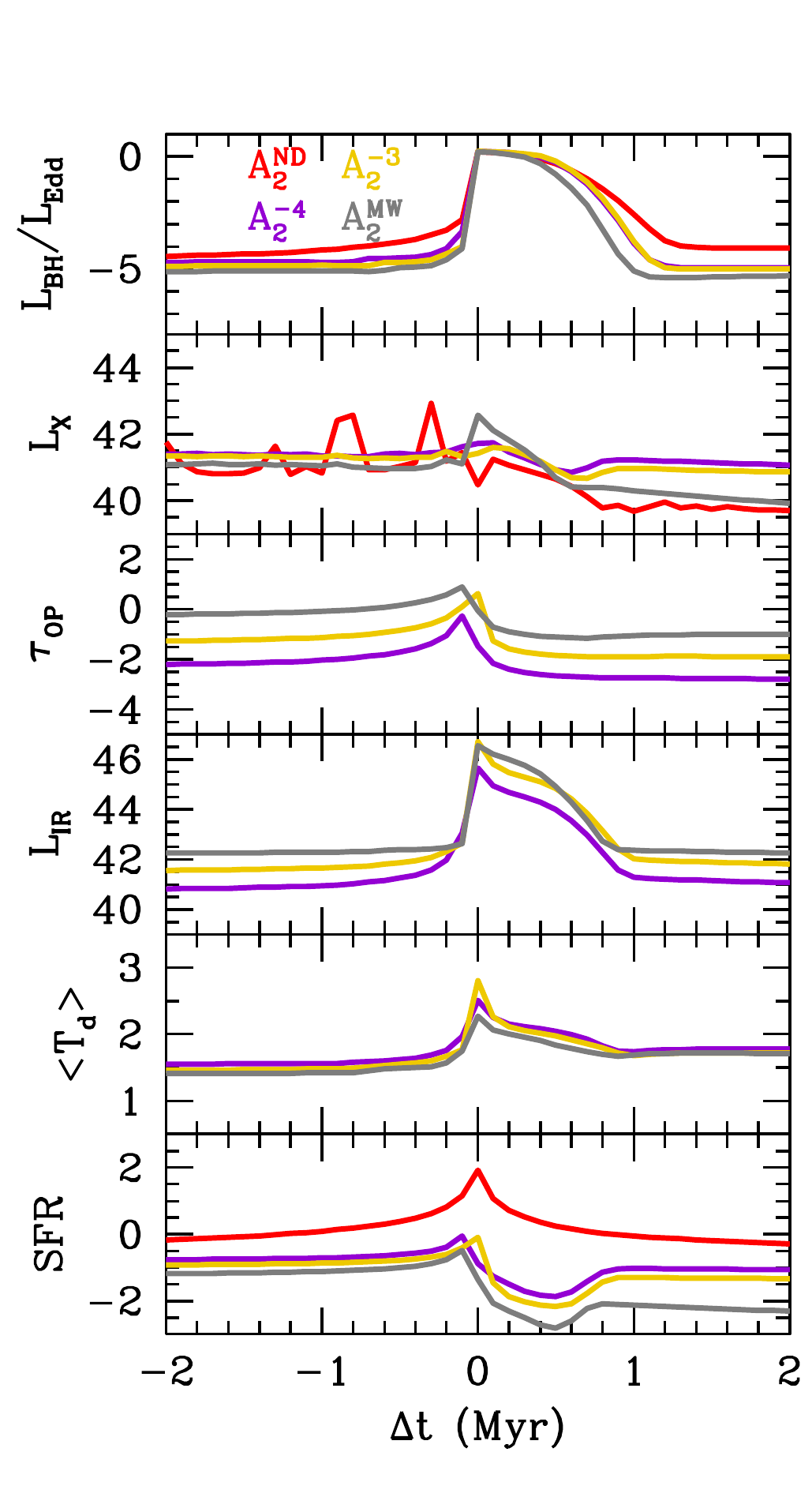}}
  \scalebox{0.6}{\includegraphics[width=14.0cm,angle=0,
                 clip=true,trim=1.0cm 1.0cm 8.8cm 0.0cm]{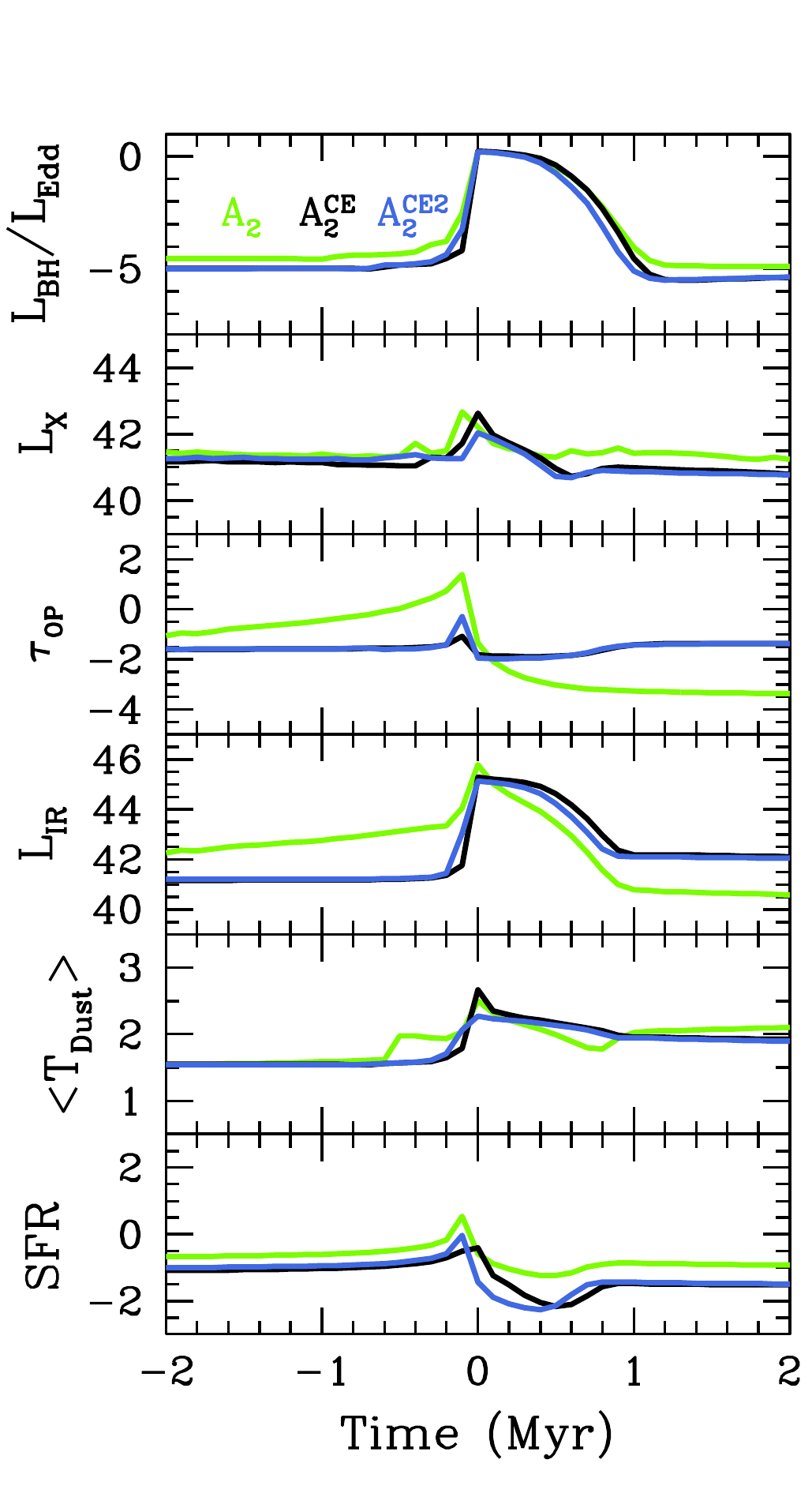}}
  \caption{\emph{Left Panels}: Time evolution of relevant quantities
    during a burst in models with constant dust to gas ratios. For
    each model, a burst was selected near 3 Gyr and scaled such that
    the maximum $L_{\rm BH}/L_{\rm Edd}$ occurs at $\Delta t = 0$. We
    plot $A_2^{\rm ND}$ in red, $A_2^{-4}$ in violet, $A_2^{-3}$ in
    gold, and $A_2^{\rm MW}$ in gray. From top to bottom, the panels
    give the X-ray luminosity in the 0.38 keV band from the hot
    emitting ISM in erg/s, the optical depth to the center of the galaxy in the
    optical band, the dust luminosity in erg/s, the luminosity-weighted dust
    temperature (see Equation~\ref{eq:tdwgt}) in K, and the star
    formation rate in $M_\odot$/yr. All y-axis quantities are given as
    $\log_{10}$. Note that the $A_2^{\rm ND}$ model
    is not plotted in the $\tau_{\rm Op}$, $L_{\rm IR}$, and $<T_d>$
    panels since it has no dust and thus a value of zero for each of
    these quantities. \emph{Right Panels}: Same as Left, but for
    models with dust to gas ratios that vary with time and
    radius. $A_2$ is plotted in green, $A_2^{\rm CE}$ in black, and
    $A_2^{\rm CE2}$ in blue.} \label{fig:burst} 
\end{figure*}

Figure~\ref{fig:burst} demonstrates the effects of changing the dust
content of the gas on the burst dynamics. One burst episode was
selected from each model near 3 Gyr, then scaled such that the maximum
$L_{\rm BH}/L_{\rm Edd}$ occurs at $\Delta t = 0$. We note that the
short few Myr duration of the bursts in these models are a feature of the A
family of models, and that the more complicated B models have burst
durations of $\simeq10$ Myr. While the common epoch for the burst
ensures some level of consistency in the galaxy evolution among
models, the $A_2^{\rm ND}$ model has undergone significantly more
black hole growth by this time than the other models, which must be
taken into account when interpreting the results. For each model, we plot a number of
relevant quantities that change through the burst-- the X-ray luminosity of the hot ISM in the 0.38
keV band, the optical depth to the center of the galaxy in the optical
band, the dust luminosity $L_{\rm IR}$, $<T_d>$ as described in
\updated{Equation}~\ref{eq:tdwgt}, and the SFR.

Overall, the burst evolution
follows the qualitative picture of the hydrodynamics given at the
beginning of this section irrespective of dust treatment. The black
hole luminosity rises rapidly followed by a decline due to the
expansion of gas in the central region. Coincident with the peak in
black hole luminosity are peaks in both $L_{\rm IR}$ and the star
formation rate (SFR). Following this positive feedback on star
formation, the SFR drops due to the AGN feedback.

There are small but important differences in the evolution among the
models. The trends are best illustrated by the models with constant
dust to gas ratios, as the changes are often monotonic with this
ratio. For instance, the dustier models have a faster decline in black
hole luminosity after the peak. The black hole is more effective in
pushing gas away in models with more dust, which slows
accretion. Similarly, the drop in SFR is monotonic in dust to gas
ratio since feedback is faster and more effective in dustier
models.

%%%%%%%%%%%%%%%%%%%%%%%%%%%%%%%%%%%%%%%%%%%%%%%%%%%%%%%%%%%%%%%%%%%%%%%%%%%
% Figure-- Burst: Constant D/G Models
%%%%%%%%%%%%%%%%%%%%%%%%%%%%%%%%%%%%%%%%%%%%%%%%%%%%%%%%%%%%%%%%%%%%%%%%%%%

We now ask how this picture is modified when the dust abundance is
treated in a more realistic way. The simplest physical treatment is
the $A_2$ model where the dust depletion is a simple function of
temperature. The bursting behavior  of this model is illustrated in the
right-hand side of 
Figure~\ref{fig:burst}. Prior to the burst, this model looks very much
like the $A_2^{\rm MW}$ model since it has high values of $\tau_{\rm
  Op}$ and $L_{\rm IR}$. After the burst, however, the AGN heats the
gas in the galaxy, which, due to the temperature-dependent dust to
gas ratio, instantaneously destroys the dust. Indeed, the $A_2$ model
closely resembles the $A_2^{\rm ND}$ model following the peak, notably
in the slightly enhanced duration of the burst and its relative lack
of suppressed star formation.

In the $A_2^{\rm CE}$ and $A_2^{\rm CE2}$ models, the dust must form and be destroyed on more
realistic timescales. In the cold shells, the
decreased temperatures allow for grain growth via collisions. From
Figure~\ref{fig:nu_ratio}, the grain growth time in a $10^4$ K shell
of cold gas with density $10^3$ cm$^{-3}$ is $\simeq$0.01 Myr, short
enough to ensure the shell is dusty. However, if the shells do not
reach these densities, the growth time can become long compared to the
infall time, rendering the dust unable to affect the dynamics. This is in contrast to
$A_2$ in which cold gas would by assumption be immediately
restored to MW-like grain abundances. Indeed, the right panel of
Figure~\ref{fig:burst} has little evidence for enhanced dust for
either the $A_2^{\rm CE}$ or $A_2^{\rm CE2}$ models, nor does the total dust mass plotted in Figure~\ref{fig:dustmass}
show any evidence for enhancement during prior to $\simeq$5 Gyr despite many
bursts. However, the existing dust is able to affect
the dynamics in ways comparable to the $A_2^{-3}$ and $A_2^{-4}$
models, notably more effective AGN feedback leading to shorter bursts
and suppressed star formation. The mixing time in the $A_2^{\rm CE2}$
model does not appear to have noticeable effects on the hydrodynamics
on the timescale of this burst.

By inspection of Figures~\ref{fig:lumconst} and \ref{fig:lumvar}, it
is obvious that not all bursts are as sharp as that expanded in
Figure~\ref{fig:burst}. In general, a series of bursts culminates in a
stronger final burst with considerably more time structure. These
episodes are easily identified in Figures~\ref{fig:lumconst} and
\ref{fig:lumvar} as the thickest bands. We recall that in the B family
of models, not discussed in this paper, the majority of bursts are of
this kind. These bursts have longer duration during which a
significant amount of material is accreted and are usually followed
by long periods of quiescence for the galaxy. Models with little or no
dust have correspondingly less radiative feedback from the AGN, and
are thus characterized by short, clean bursts. In contrast, the
dustier models have more long duration bursts as more
material is allowed to build up and then accrete. 

\begin{figure*}[htp]
 \centering
  \scalebox{0.62}{\includegraphics[width=14.0cm,angle=0,
                 clip=true,trim=0.0cm 1.0cm 5.0cm 0.0cm]{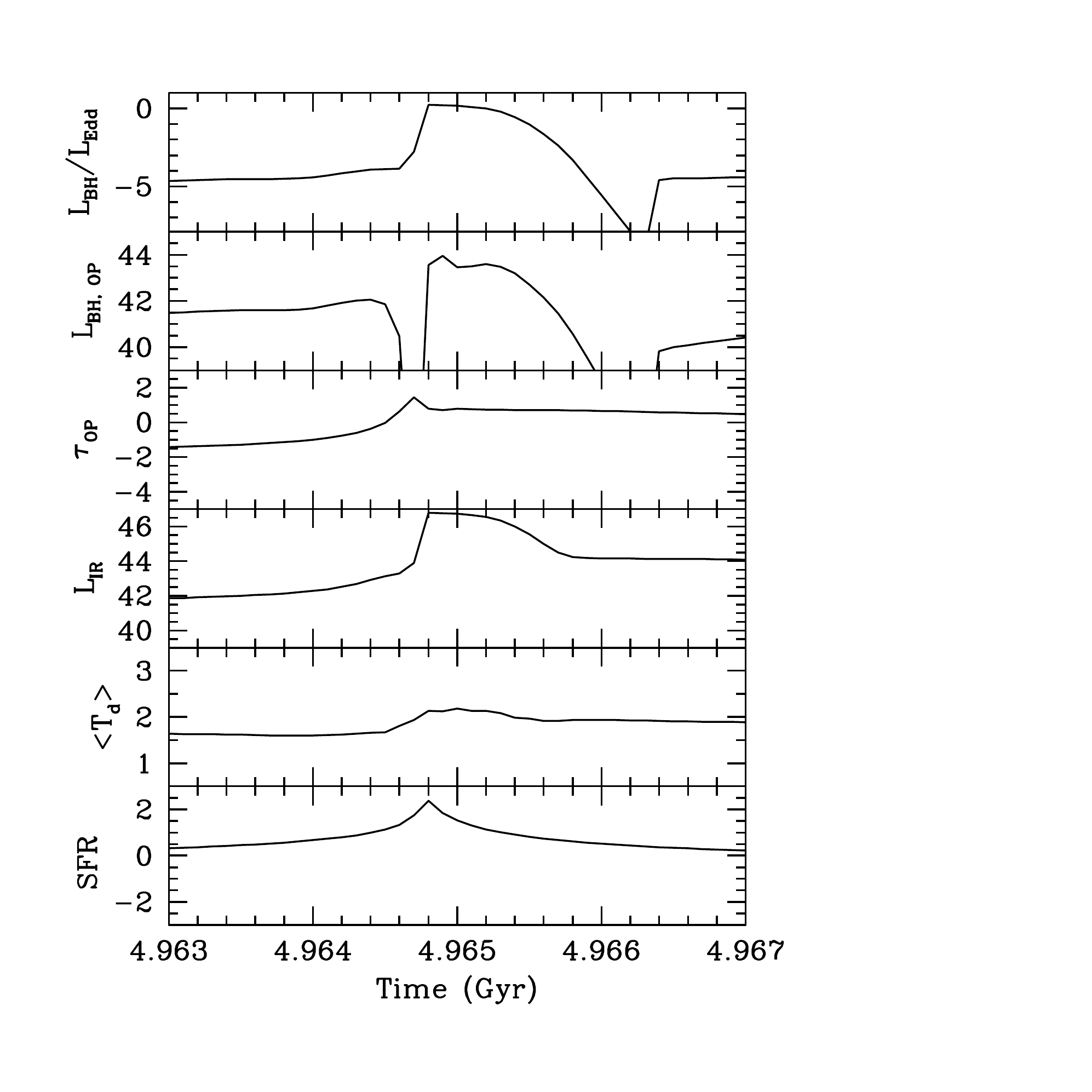}}
  \scalebox{0.62}{\includegraphics[width=14.0cm,angle=0,
                 clip=true,trim=0.0cm 1.0cm 5.0cm 0.0cm]{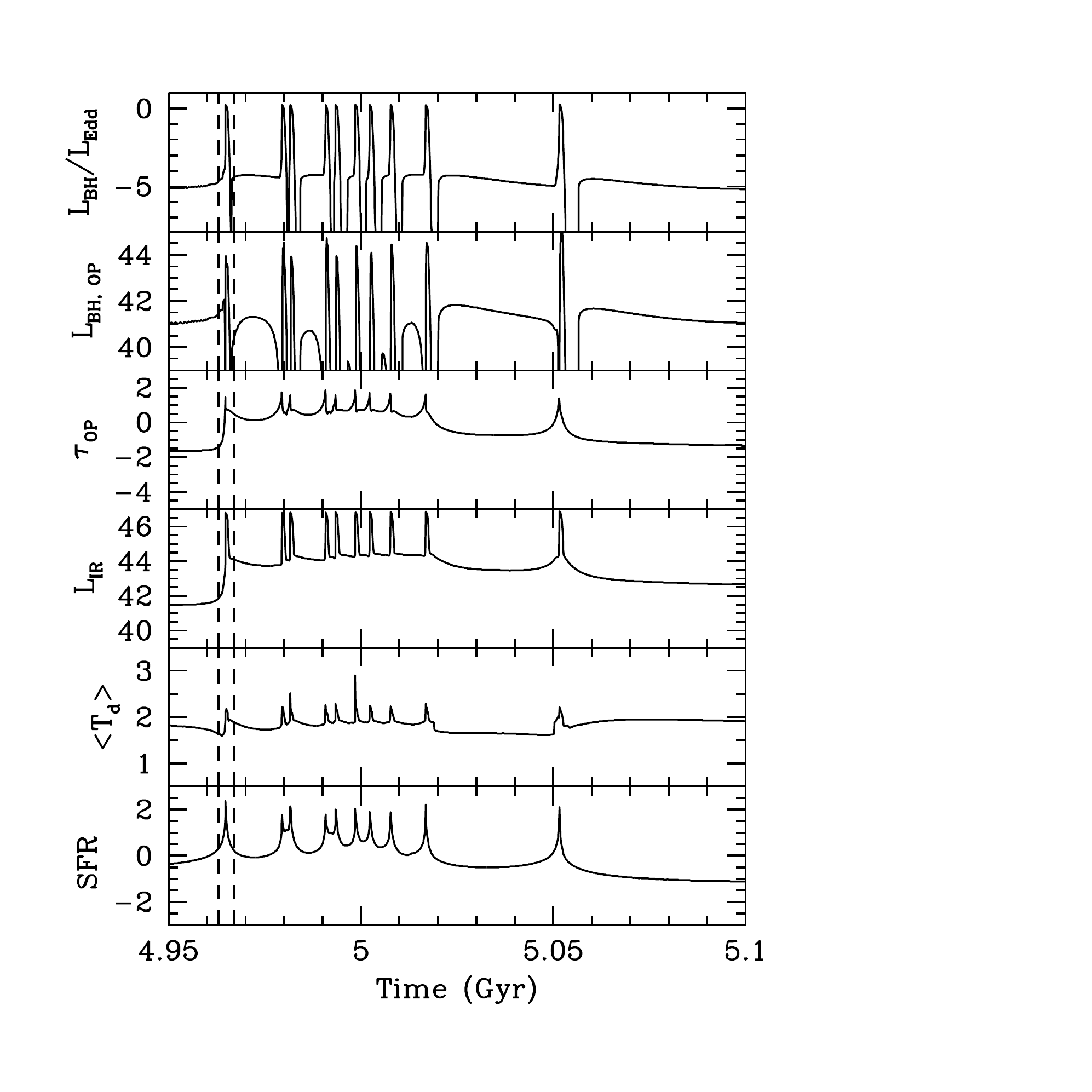}}
\caption{A major accretion event in the $A_2^{\rm CE}$ model. We
  present the first 4 Myr (the same time scale as
  Figure~\ref{fig:burst}) of this burst in the left panel, and 150
  Myr evolution of this burst in the right panel with the time limits
  of the right panel indicated by dashed lines. From top to bottom,
  we give the Eddington fraction, the effective optical luminosity of
  the black hole, the optical depth in the optical band, the IR
  luminosity from radiation reprocessed by dust, the emission-weighted
  temperature, and the star formation rate in $M_\odot$/yr. All y-axis
  quantities are given as $\log_{10}$.} \label{fig:ce_burst} 
\end{figure*}

For illustration, in Figure~\ref{fig:ce_burst} we
expand the burst around 5 Gyr in model $A_2^{\rm CE}$ and give the
Eddington fraction, effective optical luminosity from the black hole,
$\tau_{\rm Op}$, $L_{\rm IR}$, $<T_d>$, and the SFR through the
burst. Note that the time axis in the left panel spans 150 Myr,
while the right panel has the same 4 Myr span as in
Figure~\ref{fig:burst}. 

\begin{figure*}[htp]
 \centering
  \scalebox{0.62}{\includegraphics[width=14.0cm,angle=0,
                 clip=true,trim=0.0cm 1.0cm 6.0cm 0.0cm]{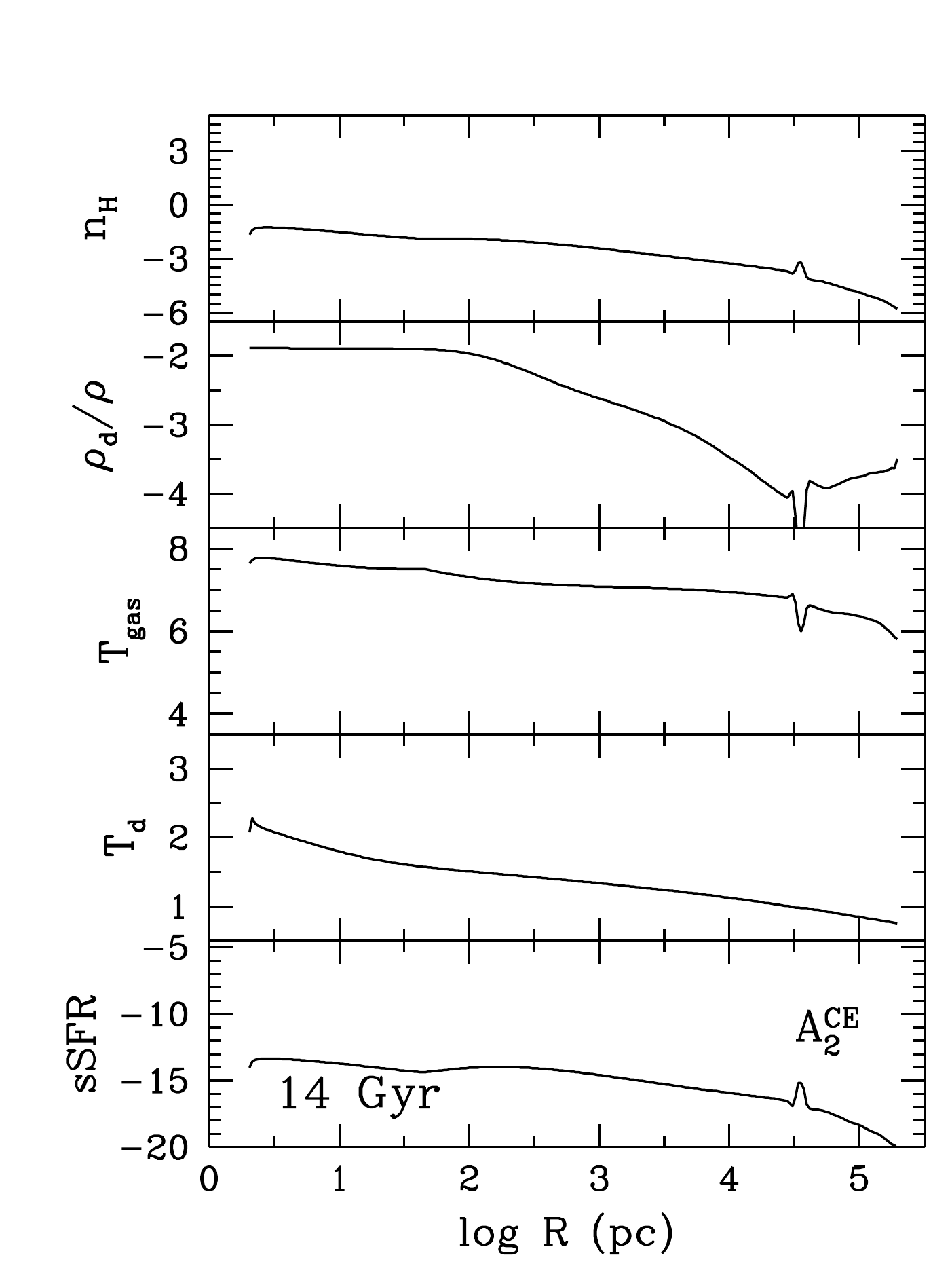}}
  \scalebox{0.62}{\includegraphics[width=14.0cm,angle=0,
                 clip=true,trim=0.0cm 1.0cm 6.0cm 0.0cm]{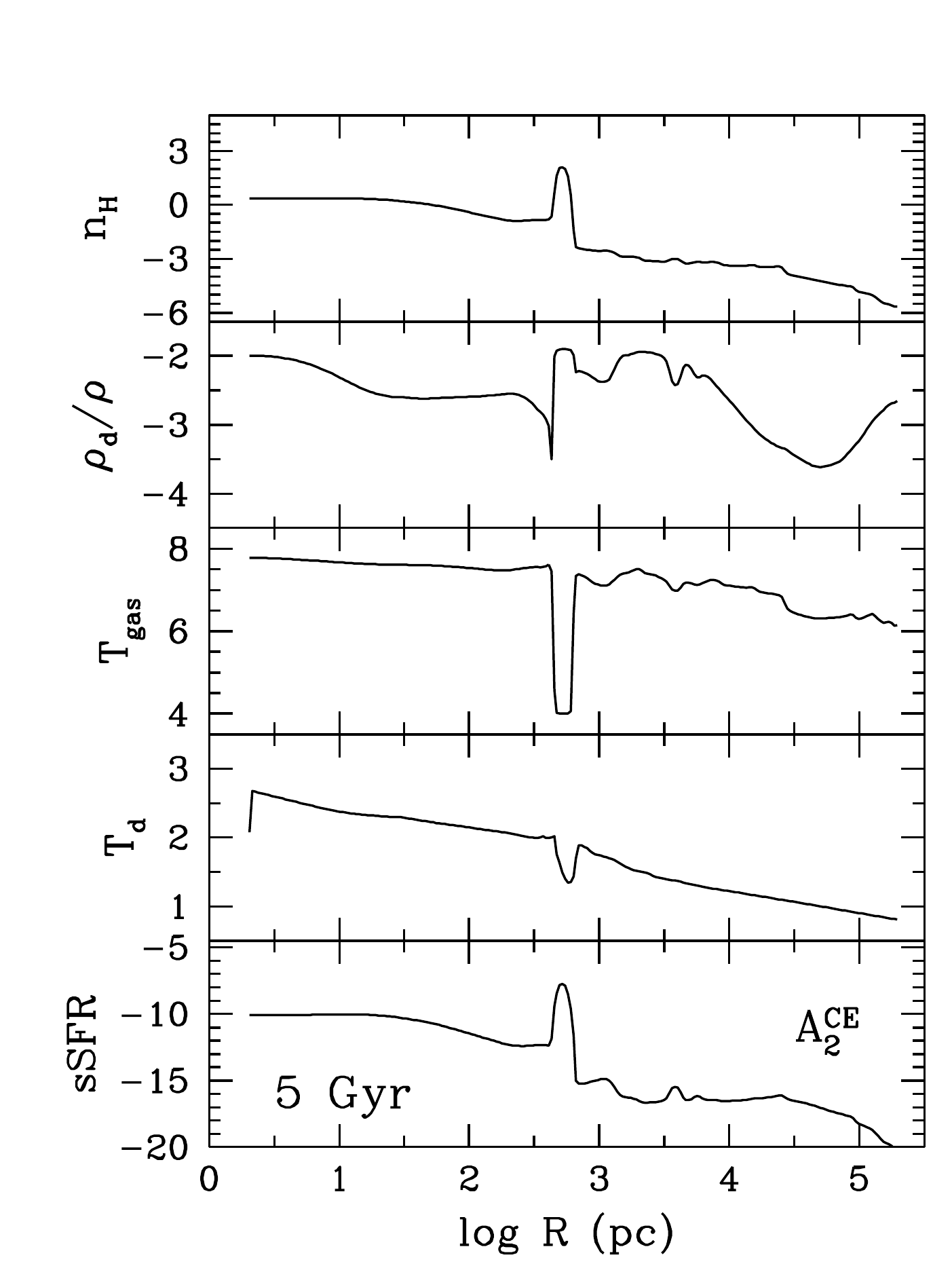}}
\caption{The radial profile of the $A_2^{\rm CE}$ model at 5 and 14
  Gyr. At 5 Gyr, the galaxy is in a prolonged period of high optical
  depth and bursting activity (see Figure~\ref{fig:ce_burst}), while
  at 14 Gyr it is quiescent. The panels are organized as follows, from
top to bottom: the gas number density in cm$^{-3}$; the dust to gas
ratio; the gas temperature in K; the dust temperature in K; the
specific star formation rate in M$_\odot$/yr/pc$^3$.  All quantities
are given as log$_{10}$. The presence of a
cold dense shell is evident at $\simeq500$ pc. Due to the star
formation activity, the dust abundance is relatively high in the inner parts of
the galaxy at 5 Gyr. Even at 14 Gyr, the galaxy maintains a high dust
to gas ratio in the inner $100$ pc.} \label{fig:ce_rprof} 
\end{figure*}

In this model,
the dust optical depth has sufficient time to rise above unity
before being stopped by the dust-destroying AGN luminosity due to the
buildup of a large, dense shell. Once the large
structure is able to collapse, a cavity
forms as the newly-fueled central AGN is able to drive out gas, dropping the accretion rate to effectively zero. In this
time, gas will again accumulate until it becomes cool and dense enough
to accrete. The galaxy oscillates between these modes for tens of Myr before
returning to the equilibrium configuration, as illustrated in the left
panel of Figure~\ref{fig:ce_burst}. Due to the prolonged
existence and extreme density of the cold shell, grain growth becomes important, with
spikes of grain growth evident in Figure~\ref{fig:dustmass} and the
dust to gas ratio saturated in the cold shell evident in Figure~\ref{fig:ce_rprof}. These
dramatic bursts illustrate the close interplay between grain growth
in cold shells and the radiation pressure that supports them.

%%%%%%%%%%%%%%%%%%%%%%%%%%%%%%%%%%%%%%%%%%%%%%%%%%%%%%%%%%%%%%%%%%%%%%%%%%%
% Figure-- Continuity Burst:
%%%%%%%%%%%%%%%%%%%%%%%%%%%%%%%%%%%%%%%%%%%%%%%%%%%%%%%%%%%%%%%%%%%%%%%%%%%

%%%%%%%%%%%%%%%%%%%%%%%%%%%%%%%%%%%%%%%%%%%%%%%%%%%%%%%%%%%%%%%%%%%%%%%%%%%
% Figure-- Continuity Radial Profile:
%%%%%%%%%%%%%%%%%%%%%%%%%%%%%%%%%%%%%%%%%%%%%%%%%%%%%%%%%%%%%%%%%%%%%%%%%%%

While the two models employing the dust continuity equation have
similar overall behavior, the ``two stream'' approach results in more
burst\updated{s} of this nature. This can be attributed to the lag time between
dust creation and mixing, which makes the AGN feedback less effective
in the early stages of the burst before mixing can occur.

\section{Observational Properties of Models}
\label{sec:observations}

Each simulation discussed above can be assessed by its ability to
reproduce the observed properties of elliptical galaxies containing
supermassive black holes. Additionally, we can assess the importance of
dust in the determination of each of the observational characteristics
we present by analyzing the variation in these quantities among the
simulations.

\updated{Of course, 1D simulations cannot
  adequately describe the observational signature of a galaxy viewed,
  e.g., along the jet axis. However, for most orientations, a 1D
  approach is sufficient to model the gas and dust intercepted by the
  line of sight. Comparisons to actual observations must be made
  with care bearing these limitations in mind.}

\subsection{Dust in Quiescent ETGs}
\label{subsec:etgdust}

It is well-established that early type galaxies harbor very little
dust due to rapid sputtering of grains in hot gas, with \citet{Clemens+Jones+Bressan+etal_2010} putting an upper limit on grain
lifetimes of $46 \pm 25$ Myr. However, {\em Spitzer} and {\em Herschel} have enabled study of the
dust that {\emph is} present and are providing important clues on the origin
of that dust. Here we summarize some recent results on the dust in
elliptical galaxies and compare with our simulations.

The {\em Herschel} Virgo Cluster Survey detected dust
emission in 46 of 910 ETGs in their sample
\citep{DiSeregoAlighieri+etal_2013}, with total dust masses ranging
between $7\times10^4$ and $1.1\times10^7$ M$_\odot$. They further note
that these masses are greater than expected for a passively evolving
galaxy, and cite a potential external origin for the dust.

The {\em Herschel} Reference Survey performed a similar study on 62
ETGs, detecting dust in 31 \citep{Smith+Gomez+Eales+etal_2011}. They too find that the dust masses
exceed predictions for passively evolving galaxies after accounting
for sputtering in hot gas, and posit that the
excess dust may be the result of mergers. Both studies find a lack of
correlation between the dust mass and stellar mass, casting doubt on
the hypothesis that the dust originates solely from stellar outflows.

Using far-infrared {\em Spitzer} data,
\citet{Temi+Brighenti+Mathews_2007a} analyzed the SEDs of 46
elliptical galaxies, finding large ($\sim$100) variations in $70\
\mu$m and $160\ \mu$m luminosity even for ellipticals with the same
B-band luminosity. Six galaxies showed extended $70\
\mu$m emission that was in excess of what would be predicted by dust
production and sputtering rates. Further, none of the galaxies showed
evidence of recent mergers and indeed some had quite old stellar
populations. Observing dust emission as extended as 5 - 10 kpc, the authors suggest that the dust has been buoyantly transported
out from a dusty nuclear region on a timescale less than the
sputtering time.

\citet{Martni+Dicken+Storchi-Bergmann_2013} use {\em Spitzer}
observations of 38 ETGs
to conclude that ETGs without dust lanes tend to have
less than $10^5 M_\odot$ of dust. Additionally, like
\citet{DiSeregoAlighieri+etal_2013} and
\citet{Smith+Gomez+Eales+etal_2011}, there is a large scatter in
the inferred dust mass at a fixed stellar mass. Like
\citet{Temi+Brighenti+Mathews_2007a}, they conclude that mergers
cannot alone count for the excess dust as the expected merger rate is
too slow relative to the dust destruction time. They propose instead
that grain growth can occur in externally accreted cold gas, with the
enhanced lifetimes of the dust in the cold gas sufficient to explain
the excess.

In our continuity models, which do not include any non-secular processes such as
mergers or accretion of cold gas from the IGM, dust growth is able to
occur in the cold gas produced by cooling flow instabilities. The
presence of such gas is attested by multi-wavelength observations of
giant ellipticals, revealing a cold ISM component
\citep{Werner+etal_2013}. In the
$A_2^{\rm CE}$ model in particular, the dust to gas ratio at $z=0$ is
a typical $10^{-4}$ while the dust mass is $\simeq10^6$ M$_\odot$,
values in accord with \citep{Smith+Gomez+Eales+etal_2011}. Thus, our
most detailed model is able to reconcile observations with theoretical
estimates of dust production and destruction rates.

Additionally, the $A_2^{\rm CE}$ and $A_2^{\rm CE2}$ models predict that the distribution
of dust in a quiescent galaxy (see Figure~\ref{fig:ce_rprof}, left
panel) that is concentrated within the inner 100 pc and then sharply
declining. Although the right panel of Figure~\ref{fig:ce_rprof} is a
snapshot of the radial profile in the midst of a complex burst, the
AGN luminosity at that precise time (see Figure~\ref{fig:ce_burst}) is
very low, and thus this object too would be interpreted as
quiescent. The dust distribution of the galaxy at this time is
markedly different, with high dust to gas ratios seen out to $\simeq
10$ kpc scale, similar to what is observed by
\citet{Temi+Brighenti+Mathews_2007a, Temi+Brighenti+Mathews_2007b}.

\begin{figure}[htp]
 \centering
  \scalebox{0.62}{\includegraphics[width=14.0cm,angle=0,
                 clip=true,trim=0.5cm 1.0cm 0.5cm 0.0cm]{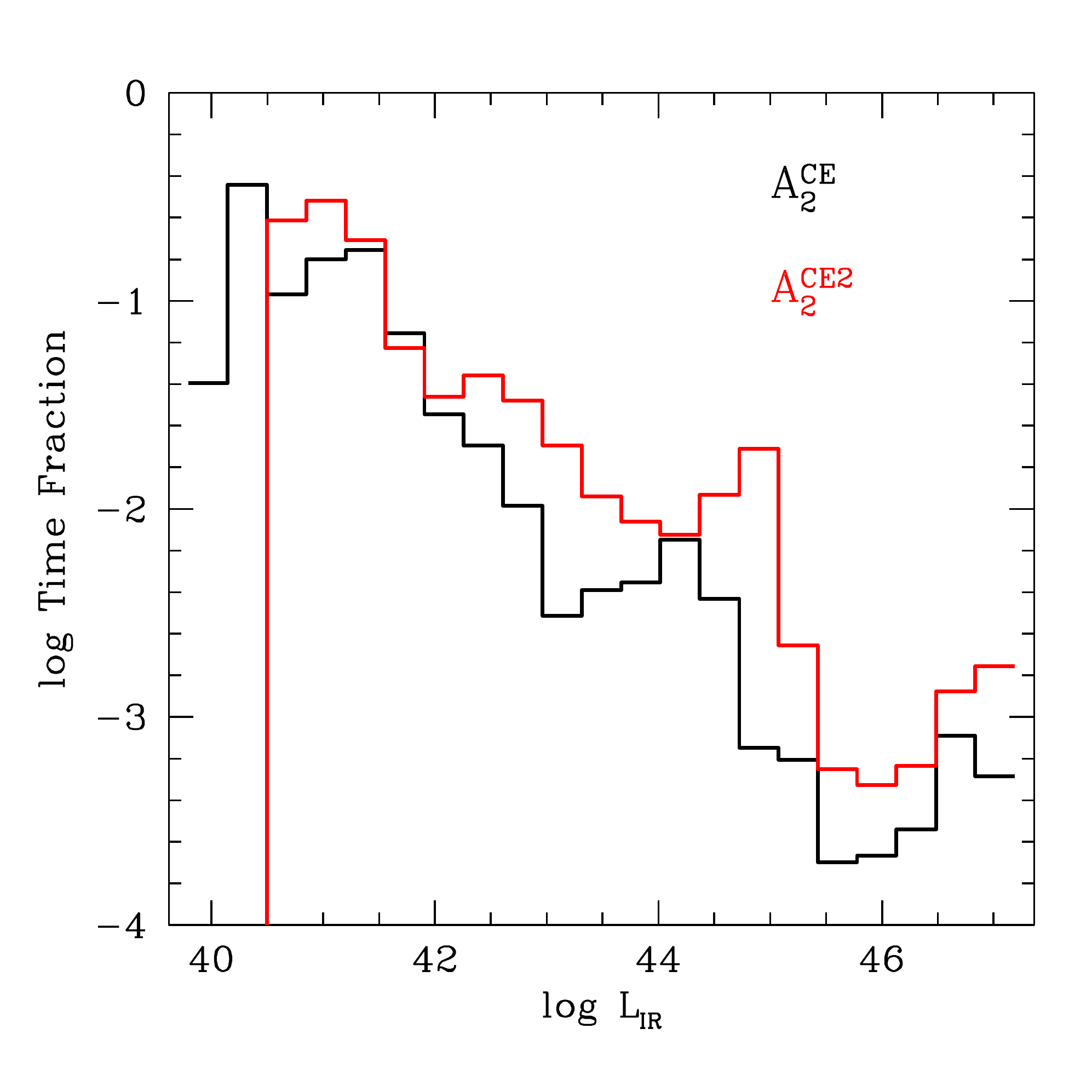}}
\caption{The IR luminosity from dust from 125,000 equally-spaced time
  slices in the simulation versus the time fraction spent in each
  luminosity bin. The majority of the time, both the
  $A_2^{\rm CE}$ (black) and $A_2^{\rm CE2}$ (red) have IR
  luminosities within a range $\simeq10^{40}-10^{42}$
  erg/s. Observations likewise indicate a large scatter in IR
  luminosity even for ellipticals at fixed stellar mass.} \label{fig:lirhist} 
\end{figure}

Finally, our models also anticipate a large variation in $L_{\rm IR}$
while $L_{\rm Op*}$ remains relatively fixed. In
Figure~\ref{fig:lirhist} we give the histogram of the infrared dust
luminosity in equally-spaced time intervals over the
simulation. In both the $A_2^{\rm CE}$ and $A_2^{\rm CE2}$ models, the
typical dust luminosity varies between $\simeq10^{40}-10^{42}$ erg/s.

%%%%%%%%%%%%%%%%%%%%%%%%%%%%%%%%%%%%%%%%%%%%%%%%%%%%%%%%%%%%%%%%%%%%%%%%%%%
% Figure-- LIR Histogram
%%%%%%%%%%%%%%%%%%%%%%%%%%%%%%%%%%%%%%%%%%%%%%%%%%%%%%%%%%%%%%%%%%%%%%%%%%%

The dust distribution predictions of our $A_2^{\rm CE}$ and $A_2^{\rm CE2}$ models lends
itself to a simple observational test. Because the dust abundance
declines sharply with radius, we find the ratio of the half radii of
the IR emission from dust and X-ray emission from the hot ISM to be
$\simeq 0.2$ in these models. In contrast, this ratio has a value of
$\simeq 1$ in models with constant dust to gas ratios.

\subsection{Dust in Galaxies with AGN}

A generic feature of all of our models is that the luminosity-weighted
dust temperature increases dramatically during bursts, usually
exceeding 100 K and often approaching the
grain sublimation temperature of $\simeq$1200 K for a brief
period. This is due to the intense AGN luminosity heating
grains in the infalling cold gas as well as the interior of the
galaxy. A key test of the viability of our models is the presence of a
significant hot dust component to the total infrared luminosity during
AGN on phases.

Using data from the AKARI Mid-Infrared Survey,
\citet{Oyabu+Ishihara_Malkan+etal_2011} discovered two LIRGs obscured in
the optical but showing strong thermal dust emission in the IR. The
derived dust temperatures were in excess of 500 K for a hot component
and ~93 K for a dominant cool component with total IR luminosity was on the
order of $10^{11} L_\odot$. Both objects were interpreted as obscured AGN, which is
broadly consistent with the predictions of our models during obscured
phases.

A key observational test of our most detailed models is the presence
of warm  dust ($T_d \simeq 100$ K) at $\leq 1$kpc during burst
events (see Figure~\ref{fig:ce_burst}, right panel). This dust is associated with the cold, dense
gas that fuels the central black hole, and while it is not close enough to the central AGN to be heated
to the sublimation temperature of grains, it is close enough to be
heated to temperatures higher than expected in the ISM of a quiescent
galaxy.

\subsection{Obscured Fraction}

\citet{Mayo+Lawrence_2013} and \citet{Lawrence+Elvis_2010} find that
only roughly 1/3 of AGN are unobscured. We assess the ``obscured
fraction'' in our models by considering how much time of the AGN-loud
phase is spent at high $\tau_{\rm Op}$. We choose the natural
threshold of $\tau_{\rm Op} > 1$ to deem the AGN ``obscured,'' which assuming a constant dust to
gas ratio of $10^{-4}$ implies a column density of $2\times10^{23}$
cm$^{-2}$ given our prescription for $\kappa_{\rm Op}$ (Equation~\ref{eq:opacity}). For the continuity model, we
find that $\tau_{\rm Op} > 1$ for $54\%$ of the time that the AGN is
on ($L_{\rm BH}/L_{\rm Edd} > 1/30$), the two-stream $73\%$, the standard $A_2$
model $44\%$, the constant $10^{-2}$ depletion model $10\%$, and the
model with MW dust abundance $3\%$. It is
clear that to obtain the observed high obscuration fractions it is necessary to
decouple the dust abundance from the gas abundance-- models with too
much dust cannot sustain accretion and high Eddington ratios while
models with little dust provide minimal obscuration. Only by allowing
the dust to be formed and destroyed in a physical way do we see the
emergence of clear obscured and unobscured phases directly related to
the ability of the AGN to drive and quench star formation, and
consequently dust production.

\subsection{Star Formation Rate}
As already described in CO07, all of our models predict a period of AGN-induced star formation,
the so-called ``positive feedback'' \citep[\updated{see also}][]{Ishibashi+Fabian+2012, Zubovas+etal_2013}, with star formation occurring within the inner few
hundred parsecs in the galaxy. Following this period, star formation
is quenched to below pre-burst levels. Though the
interplay is complex, it is evident that the black hole accretion rate (BHAR)
and the star formation rate (SFR) are closely entwined. 

\citet{Chen+etal_2013} sought evidence of a BHAR-SFR relationship by
studying the {\it average} BHARs of AGN as determined by their X-ray
luminosity and looking for correlations with the SFR as inferred from
the IR luminosity. Due to the intense variability of AGN on timescales
short compared to star formation time, averaging is emphasized as
painting a clearer picture of the relationship. They find that

\begin{equation}
\log\left(L_X{\rm [erg/s]}\right) = 30.37 + 1.05\log\left(L_{\rm
    IR}/L_\odot\right)
~~~,
\end{equation}
for their best-fit model, which analyzed galaxies in the redshift
range $0.25 < z < 0.8$ and with SFRs $0.85 <$ log SFR/M$_\odot$ $<
2.56$. Converting to BHAR and SFR, they obtain

\begin{equation}
\log {\rm BHAR} = -3.72 + 1.05\log{\rm SFR}
~~~.
\end{equation}

\begin{figure*}[htp]
 \centering
  \scalebox{0.62}{\includegraphics[width=14.0cm,angle=0,
                 clip=true,trim=0.5cm 1.0cm 0.5cm 0.0cm]{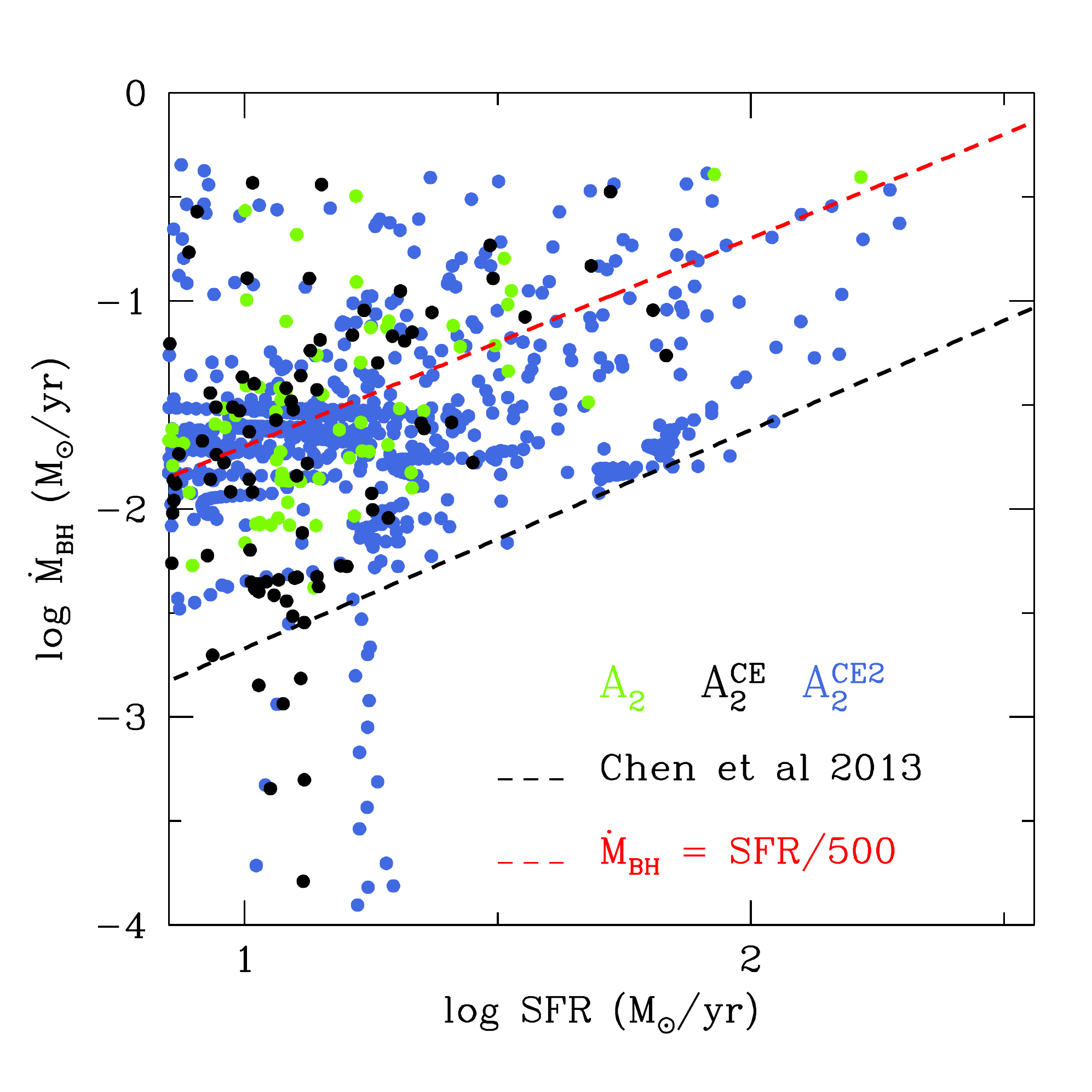}}
  \scalebox{0.62}{\includegraphics[width=14.0cm,angle=0,
                 clip=true,trim=0.5cm 1.0cm 0.5cm 0.0cm]{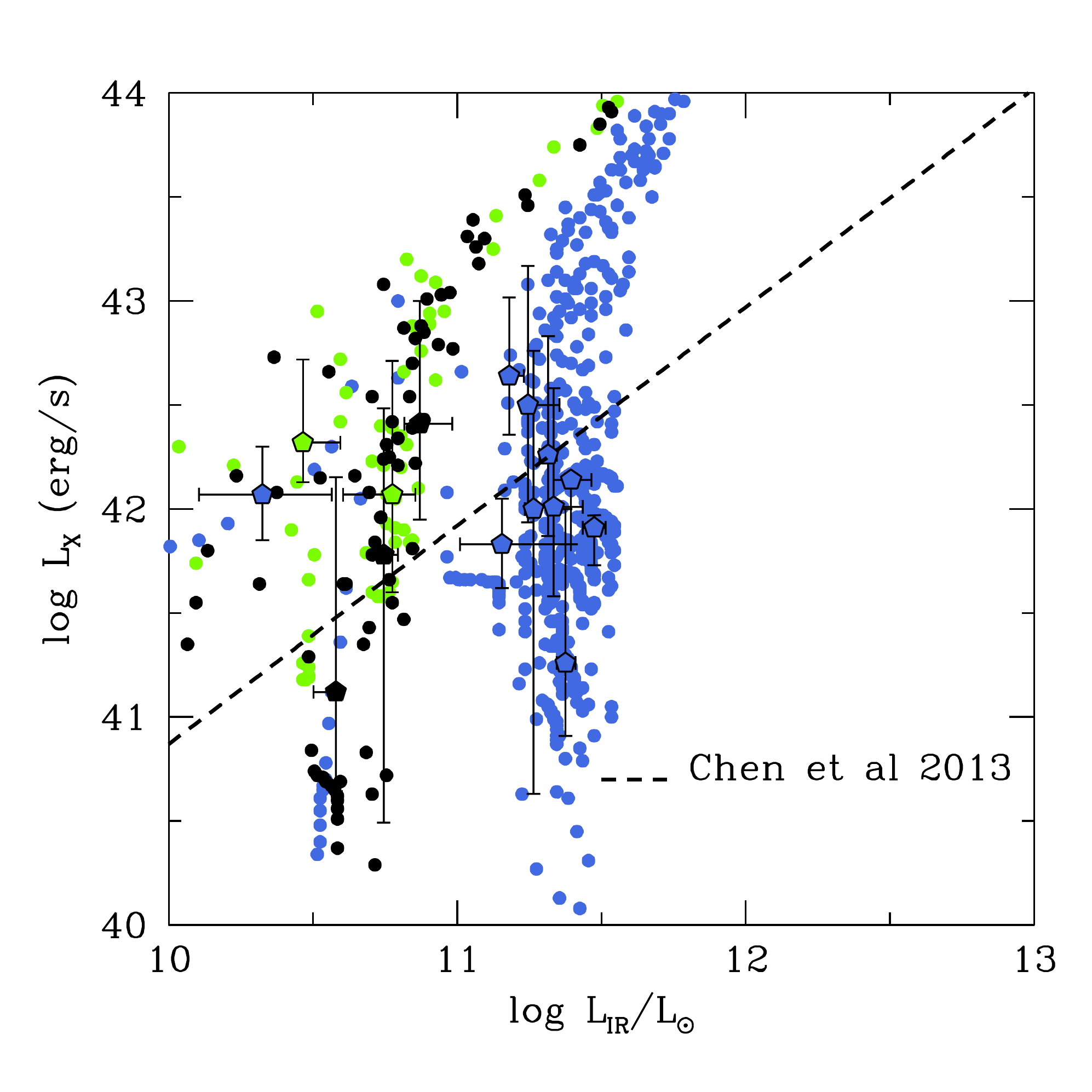}}
\caption{The correlation between the black hole accretion rate (BHAR)
  and the star formation rate (SFR) between redshifts 0.25 and
  0.8. Both observations by \citet{Chen+etal_2013} and the simulations
find a power law relationship with index of $\simeq$1. The picture is
less clear when looking at $L_X$ and $L_{\rm IR}$, the more
fundamental observables, due to variations in $L_X$ at fixed $L_{\rm
  IR}$. \updated{Using pentagonal symbols, we plot the median value
  of all points within 50 Myr time bins. The error bars
  indicate the upper and lower quartiles of each bin.}} \label{fig:chen_sfr} 
\end{figure*}

To compare this result with our most physical simulated galaxies, in
Figure~\ref{fig:chen_sfr} we
make the same cuts in redshift and SFR and consider the BHAR and SFR
in the simulation at equally-spaced times. Since the observational
data do not have objects with $L_X > 10^{44}$ erg/s, and since these
objects are likely to be obscured in our simulations at variance with
the $10^{20}$ cm$^{-2}$ column density assumed by
\citet{Chen+etal_2013}, we removed all points with $L_X > 10^{44}$
erg/s.

Indeed, there is a strong linear
correlation in all models between the BHAR and SFR. However, the
points cluster more closely to the line BHAR = SFR/500, which
\citet{Chen+etal_2013} derived from the M$_{\rm BH}$-M$_{\rm bulge}$
relations of \citet{Marconi+etal_2004} than to the observations of
\citet{Chen+etal_2013}. Nevertheless, the slopes appear
consistent. The $L_X$-$L_{\rm IR}$ plot varies significantly from
the BHAR-SFR plot for our models. This could be partially due to rapid
variations in the X-ray luminosity at relatively constant $L_{\rm
  IR}$, to which the $A_2^{\rm CE2}$ model would be particularly
susceptible given its delayed dust mixing. \updated{Averaging the data
  in 50 Myr time bins (since we cannot average over an ensemble of
  galaxies) brings the simulations into reasonable agreement with the
  observational data as shown
  in Figure~\ref{fig:chen_sfr}, although the uncertainties are large.}

It must also be
noted that our models consider only AGN-induced star formation, and
thus by neglecting star formation induced by other processes,
e.g. mergers, we are likely under-predicting the total star
formation rate. Secondly, we are modeling a single galaxy with a single
velocity dispersion, not an ensemble of galaxies, so a quantitatively
exact comparison is beyond the scope of this work. These caveats
notwithstanding, AGN-induced star formation appears at least roughly
consistent with the observed SFR-BHAR correlation.

%%%%%%%%%%%%%%%%%%%%%%%%%%%%%%%%%%%%%%%%%%%%%%%%%%%%%%%%%%%%%%%%%%%%%%%%%%%
% Figure-- BHAR vs SFR
%%%%%%%%%%%%%%%%%%%%%%%%%%%%%%%%%%%%%%%%%%%%%%%%%%%%%%%%%%%%%%%%%%%%%%%%%%%

\section{Discussion and Conclusions}
\label{sec:discussion}
By implementing a more physically-based dust treatment into 1D
hydrodynamical simulations of the evolution of massive elliptical
galaxies, we are able to link the computed IR emission from the
galaxy during various stages of secular evolution with observations of
IR emission. These models are capable of
attaining LIRG and ULIRG-like phases of high IR emission without
needing to invoke non-secular processes.

Despite the differing assumptions on dust abundance, the simulated
galaxies illustrated a remarkably robust mass budget-- in each
simulation, the vast majority of the gas in the galaxy was expelled in
outflows, about $10\%$ was turned into stars, a few percent was
accreted onto the central black hole, and a few percent remained as
gas. The black hole growth is consistent both with current determinations
of the Magorrian relation and the empirical fact that quasar ``on''
periods decline in frequency with decreasing redshift.

Our most physical dust models are able to reconcile the low observed dust
abundance of quiescent galaxies (dust to gas ratios of
$\simeq10^{-4}$) with presence of heavily obscured quasars
through grain growth and reduced sputtering rates in cold gas. Additionally,
optically-thick gas was able to oscillate between accretion and
outflow phases for tens of Myr, resulting in sustained periods of
large IR luminosity consistent with LIRGs and ULIRGs. However, at variance with
\citet{Debuhr+Quataert+Ma_2011}, $\tau_{\rm IR}$ never exceeds
unity and the momentum imparted to the dust gas never exceeds $L_{\rm
  BH}/c$. 

We identify two distinct types of AGN bursts common to all models-- short-duration
optically thin bursts that eventually culminate to a single large,
complex burst that is largely optically thick. A clear prediction of this work is the presence of infrared
emission from $\simeq100$ K dust grains in the inner $\simeq1$ kpc of
massive galaxies during AGN bursts.

The presence of dust grains in accreting gas was also found to impact
the star formation processes in the galaxy-- AGN feedback and
consequent quenching of star formation was enhanced in models with
more dust. Similarly, dusty models also accrete less gas and have
shorter duration bursts. Irrespective of our dust treatment, we find periods of ``positive
feedback'' on star formation in which AGN activity precipitates a
brief period of active star formation.

An inherent limitation of 1D simulations is the inability to account
for fragmentation of gas. The influence of dust in this case,
particularly in its role of preventing gas from accreting, has yet to
be determined using a detailed physical prescription for the dust
abundance. The formalism laid out in this work can be easily
generalized to higher dimensional simulations, and given the
importance of dust not only in the dynamics but also observational
signatures, doing so may shed additional light on evolution of these
galaxies.

\acknowledgments
{We thank Bruce Draine, Jenny Greene,  Jill Knapp, and Greg Novak for
  helpful discussions \updated{and the anonymous referee for detailed feedback
  that significantly improved the quality of this paper}. BH acknowledges support from the National Science Foundation
  Graduate Research Fellowship under Grant No. DGE-0646086. LC is
  supported by the grant MIUR 2008, and PRIN MUIR 2010-2011, project
  ``The Chemical and Dynamical Evolution of the Milky Way and Local
  Group Galaxies'' 2010LY5N2T.}

\bibliography{agn_dust}

\end{document}